\documentclass{article}

\usepackage[a4paper]{geometry}
\usepackage[utf8]{inputenc}
\usepackage[american]{babel}
\usepackage{graphicx}
\usepackage{amsmath}
\usepackage{amssymb}
\usepackage{bm}
\usepackage{xcolor}
\usepackage{xr-hyper}
\usepackage{hyperref}
\usepackage[acronym]{glossaries}
\glsdisablehyper
\usepackage{authblk}
\usepackage[square,numbers]{natbib}

\newacronym{mfpc}{MFPC}{multivariate functional principal component}
\newacronym{mjm}{MJM}{multivariate joint model}
\newacronym{ufpc}{UFPC}{univariate functional principal component}
\newacronym{mcmc}{MCMC}{Markov chain Monte Carlo}
\newacronym{mfpca}{MFPCA}{multivariate functional principal component analysis}
\newacronym[
  longplural={univariate functional principal component analyses}
]{ufpca}{UFPCA}{univariate functional principal component analysis}
\newacronym{iid}{i.i.d.}{independent and identically distributed}
\newacronym{kl}{KL}{Karhunen-Lo\`eve}
\newacronym{mse}{rMSE}{root mean squared error}
\newacronym{pbc}{PBC}{primary biliary cholangitis}
\newacronym{inl}{INL}{integrated nested Laplace}

\begin{document}

\title{Flexible joint models for multivariate longitudinal and time-to-event data using multivariate functional principal components}

\author[1]{Alexander Volkmann*}

\author[2]{Nikolaus Umlauf}

\author[1]{Sonja Greven}
\affil[1]{Chair of Statistics, School of Business and Economics, Humboldt-Universit\"{a}t zu Berlin, Germany}
\affil[2]{Department of Statistics, Faculty of Economics and Statistics, Universit\"{a}t Innsbruck, Austria}
\date{}
\setcounter{Maxaffil}{0}
\renewcommand\Affilfont{\itshape\small}

\maketitle

\begin{abstract}
The joint modeling of multiple longitudinal biomarkers together with a time-to-event outcome is a challenging modeling task of continued scientific interest. In particular, the computational complexity of high dimensional (generalized) mixed effects models often restricts the flexibility of  shared parameter joint models, even when the subject-specific marker trajectories follow highly nonlinear courses. We propose a parsimonious multivariate functional principal components representation of the shared random effects. This allows better scalability, as the dimension of the random effects does not directly increase with the number of markers, only with the chosen number of principal component basis functions used in the approximation of the random effects. The functional principal component representation additionally allows to estimate highly flexible subject-specific random trajectories without parametric assumptions. The modeled trajectories can thus be distinctly different for each biomarker. We build on the framework of flexible Bayesian additive joint models implemented in the R-package ‘bamlss’, which also supports  estimation of nonlinear covariate effects via Bayesian P-splines. The flexible yet parsimonious  functional principal components basis used in the estimation of the joint model is first estimated in a preliminary step. We validate our approach in  a simulation study and illustrate its advantages by analyzing a study on primary biliary cholangitis.
\end{abstract}

\section{Introduction}
\label{sec:Intro}

More and more medical studies are gathering information on their participants' longitudinal profiles of different biomarkers together with the time to an event of interest such as rejection of a transplanted organ \cite{goldschmidt2018immune}, onset of a disease \cite{hakola2019association}, or death \cite{murtaugh1994primary}. It is then of interest to model the longitudinal profiles together with the time-to-event data to gain a detailed understanding of the association of the different processes and to improve prediction for new subjects.\cite{faucett1996simultaneously, wulfsohn1997joint, proust2009development} Shared parameter models are a popular approach to jointly modeling the longitudinal and time-to-event processes.\cite{papageorgiou2019overview} They typically combine mixed effects models for the longitudinal data with a hazard model for the time-to-event data through shared random effects. For multiple longitudinal outcomes, however, the generalized mixed models and thus the shared parameter models can become quite computationally expensive due to a large number of parameters to estimate. This can restrict the flexibility with which the longitudinal outcomes are modeled, even though the respective longitudinal trajectories often exhibit strong non-linearities in practice.\cite{papageorgiou2019overview, kohler2017flexible, li2022joint} Recent research has thus focused on lowering the computational costs of these \glspl{mjm} by adapting two-stage methods (fitting the longitudinal submodel and using its output to fit the time-to-event submodel)\cite{mauff2020joint, mauff2021pairwise}, improving numerical efficiency\cite{philipson2020faster}, or using approximate expectation-maximization algorithms\cite{murray2022fast, murray2023fast} or approximate Bayesian methods\cite{rustand2023fast, tu2023gaussian}.
 
Here, we propose a further means of reducing the computational complexity of \glspl{mjm}, which can be combined with the aforementioned approaches. By understanding the longitudinal profiles as functions over the follow-up time, we introduce \gls{mfpca} as a popular tool for dimension reduction in functional data analysis to the \gls{mjm} framework. This allows us to represent the shared random effects using \glspl{mfpc}, which has several advantages: An \glspl{mfpc} basis representation allows to model highly flexible functional random effects while being parsimonious at the same time. These basis functions efficiently capture correlated nonlinear trends over the different longitudinal outcomes, thus admitting a flexible modeling approach with relatively few parameters, especially compared to a (penalized) spline approach for each longitudinal outcome. As the \gls{mfpca} decomposes the entire multivariate random process, the number of \gls{mfpc} based random effects needed for a good model fit does not directly depend on the number of longitudinal outcomes but only on the complexity of the covariance structure. Furthermore, the \glspl{mfpc} form an orthogonal basis with uncorrelated random weights (due to the Karhunen-Loève theorem), which simplifies the generally unstructured random effects covariance matrix of e.g.\ random spline basis coefficients to a diagonal matrix. Finally, the \glspl{mfpc} can be ordered by the amount of variation they explain, which provides a handy criterion for further dimension reduction by dropping relatively unimportant modes of variation from the estimation and thus reducing the number of parameters in the model.

There is a wide range of work that has combined functional data analysis with joint models of (multivariate) longitudinal and time-to-event outcomes. Some of this work has focused on incorporating functional data such as images as baseline covariates or as a repeatedly measured longitudinal functional outcome.\cite{ye2015joint, li2017functional, kong2018flcrm, li2019bayesian, zou2021bayesian, kang2023joint, zou2023multivariate} In this context, the functional domain is different from the follow-up time and functional principal component analysis is used to achieve a low dimensional representation of the functional data. Here, we propose to view repeated scalar measurements as functional data so that the functional domain coincides with the follow-up time. This has also been considered by other authors, particularly in the context of two-stage models and dynamic predictions.\cite{holte2012efficient, yan2017dynamic, yan2018functional, li2019dynamic, hong2021dynamic, lin2021functional, dong2023jointly} These approaches use \gls{ufpca} or \gls{mfpca} to estimate principal components and scores for the longitudinal submodel and plug the scores into the survival submodel. While two-stage approaches are known to yield biased parameter estimates\cite{tsiatis2004joint, rizopoulos2012joint}, the previously mentioned works show that this estimation bias does not necessarily negatively affect the predictive performance. Other approaches simultaneously estimate the principal components and scores within the univariate\cite{yao2007functional} or multivariate joint model.\cite{li2021flexible, li2022joint} For multiple longitudinal outcomes, however, these proposed \glspl{mjm} feature strong assumptions. Li et al. (2021)\cite{li2021flexible} use \gls{ufpca} to model a common latent trajectory shared by all longitudinal outcomes, which allows to represent only highly correlated markers. Li et al. (2022)\cite{li2022joint} and Zou et al. (2023)\cite{zou2023bayesian} extend this approach by further including outcome specific latent trajectories, but assume them to be \gls{iid} across outcomes and subjects. In both approaches, the dependence between longitudinal outcomes is explained by only one shared latent factor, which might not be adequate for multiple biomarkers from potentially different biological pathways.

While our proposed estimation approach also involves two steps, it is important to note that this differs greatly from the previously mentioned two-stage procedures. In the preliminary step, we estimate \glspl{mfpc} via \gls{mfpca}. These are then used as an empirical, flexible, and parsimonious basis to model the multiple longitudinal outcomes within an \gls{mjm}, thus incorporating the time-to-event information together with the longitudinal modelling and avoiding estimation bias induced by the marginal two-stage approach. Volkmann et al. (2023)\cite{volkmann2023multivariate} show that this two step estimation process using estimated \glspl{mfpc} works well and provides the flexibility to fit complex multivariate functional data. We are thus extending their work to the scenario where the multivariate functional data are subject to informative missingness.

Our implementation combines the flexibility and parsimoniousness of using an \gls{mfpc} basis representation of functional random effects with the framework of flexible additive joint models proposed by K\"ohler et al. (2017)\cite{kohler2017flexible}. This framework incorporates the shared parameter models into Bayesian additive models for location, scale, and shape\cite{umlauf2018bamlss}, and provides versatile, powerful, and easy to use modeling options for covariate effects. The implementation allows to include for example smooth, linear, spatial, time-varying, or random effects terms of one or more covariates in the longitudinal and time-to-event submodels, using the well developed and documented  architecture of the \textsf{R} package \textbf{mgcv}\cite{pkg:mgcv} for generalized additive models\cite{wood2017generalized}. This supports more (and arguably easier to use) options for the specification of nonlinear covariate effects than other available \textsf{R} software\cite{hickey2018joinerml, Goodrich2020rstanarm, Murray2023gmvjoint}, in particular the popular \textbf{JMbayes2} package\cite{pkg:JMbayes2}. We adopt a derivative based \gls{mcmc} algorithm to estimate the full \gls{mjm}, thus avoiding the different proposed approximations such as via two-stage approaches as in Mauff et al. (2020)\cite{mauff2020joint}.

We illustrate our proposed approach by analyzing a well known and publicly available \gls{pbc} data set\cite{rizopoulos2012joint}. It is based on a study conducted by the Mayo Clinic from January 1974 to May 1984 on the survival of patients with \gls{pbc} (previously called primary biliary cirrhosis), which is a rare chronic liver disease causing the inflammation and destruction of small bile ducts.\cite{murtaugh1994primary} This data set is still widely used in recent literature \cite{andrinopoulou2016bayesian, murray2022fast, murray2023fast, pandolfi2023hidden, rustand2023fast, tu2023gaussian}, mainly due to two reasons: it is one of the last studies to allow the analysis of the natural history of \gls{pbc} patients treated only with supportive care \cite{fleming2013counting} and it features an interesting data structure. Several biochemical measurements are available for 312 subjects at different time points after entering the study, with missing observations and dropout due to censoring, liver transplantation, or death. Furthermore, these longitudinal measurements have previously been found to follow non-linear trajectories and to be associated with the patients' survival.\cite{andrinopoulou2016bayesian, kohler2018nonlinear, rustand2023fast} Using an \glspl{mfpc} based representation of the random effects can lower the computational burden while increasing the flexibility in the modeling of the longitudinal trajectories, which can lead to a more accurate estimation of the association between the longitudinal and the event processes.

This paper is structured as follows. Section \ref{sec:Methodology} formally introduces the flexible \gls{mjm} as well as the \gls{mfpc} representation of the random effects, while Section \ref{sec:Estimation} presents the proposed estimation of the \gls{mfpc} basis and the flexible \gls{mjm}. In Section \ref{sec:Simulations} we examine our approach in different simulation scenarios and illustrate its applicability using the \gls{pbc} data in Section \ref{sec:Application}. Section \ref{sec:Discussion} closes with a discussion and outlook.

\section{Methodology}
\label{sec:Methodology}

In the following, we present a flexible joint model for multivariate longitudinal outcomes and time-to-event data. We follow and extend the notation of K\"ohler et al. (2017)\cite{kohler2017flexible} to emphasize the high flexibility of all model parts as structured additive predictors, including modeling nonlinear covariate effects using Bayesian P-splines. We then introduce the idea of representing the shared random effects of the joint model using multivariate functional principal components, which gives a parsimonious representation of a possibly very complex covariance structure.

\subsection{Multivariate joint model}

Let $T_i \in \mathcal{I}$ denote the potentially right-censored event time for subject $i=1,...,n$ with $\mathcal{I} = [0, T_{max}]$ and $T_{max}$ the maximal follow-up time. We model the hazard $h_i(t)$ of an event for subject $i$ at time $t \in \mathcal{I}$ as a function of $K$ longitudinally observed markers and further covariates as

\begin{equation}
\label{eq:hazard_one_observation}
h_i(t) = \text{exp}\{\eta_i(t)\} = \text{exp}\{\eta_{\lambda i}(t) + \eta_{\gamma i} + \sum_{k=1}^{K} \eta_{\alpha_ki}(t)\cdot\eta_{\mu_ki}(t)\},
\end{equation}
where $\eta_i(t)$ is a structured additive predictor for subject $i$ at time $t$ given a set of baseline and time-dependent covariates $\bm{x}_i$. Note that 
 we suppress the dependency on $t$ for the set of covariates for better readability.
The additive predictor $\eta_i(t)$ can be further separated into structured additive predictors $\eta_{\lambda i}(t)$ for the baseline hazard, time-dependent covariates, and time-dependent covariate effects, $\eta_{\gamma i}$ for the baseline covariates, and the possibly time-dependent association modeled by $\eta_{\alpha _ki}(t)$ of the hazard with the $K$ longitudinal outcomes $\eta_{\mu_ki}(t)$, $k = 1,...,K$. We consider the ``true'' underlying longitudinal outcomes $\eta_{\mu_ki}(t)$ as structured additive functions over $\mathcal{I}$ and assume them to be square-integrable, $\eta_{\mu_ki}(t) \in L^2(\mathcal{I})$. These functions, however, are only observed at subject and outcome specific time-points $t_{ijk}, j = 1, ..., n_{ik}$ with corresponding longitudinal observations $y_{ijk}$ modeled as
\begin{equation}
\label{eq:longi_model_one_obs}
y_{ijk} = \eta_{\mu_ki}(t_{ijk}) + \epsilon_{ijk},
\end{equation}
where $\epsilon_{ijk}$ represents independent Gaussian noise
\begin{equation}
\label{eq:longi_model_assumpt}
\epsilon_{ijk}\overset{\text{ind}}{\sim} N(0, \exp\{\eta_{\sigma_ki}(t_{ijk})\}^2),
\end{equation}
whose variance depends on the structured additive predictor $\eta_{\sigma_k i}(t_{ijk})$.

The association predictors $\eta_{\alpha_k i}(t)$ link the multivariate longitudinal submodel of equations \eqref{eq:longi_model_one_obs} and \eqref{eq:longi_model_assumpt} to the time-to-event submodel \eqref{eq:hazard_one_observation}. The resulting \gls{mjm} connects the current outcome levels to the risk of experiencing the event at time $t$. 
Note that we estimate separate association predictors $\eta_{\alpha_k i}(t)$ for the different longitudinal predictors $\eta_{\mu_k i}(t)$, allowing us to identify longitudinal outcomes important for the time-to-event process.

The additive predictors 
\begin{equation*}
 \eta_{li}(t) = \sum_{h=1}^{H_l}f_{lh}(\bm{x}_i,t), l \in \mathcal{L} := \{\lambda, \gamma, \alpha_1,...\alpha_K, \mu_1,...,\mu_K, \sigma_1,...,\sigma_K\}
\end{equation*}
comprise parametric or nonparametric functions of $\bm{x}_i$ as well as random effects, all potentially varying over time $t$. In particular, the effects $f_h(\bm{x}_i,t)$ depend on subsets of covariates in $\bm{x}_i$; they can specify linear or smooth covariate effects of single covariates in $\bm{x}_i$ or interaction effects of two or more such covariates. As an illustration, a scalar, time-constant covariate $z_i \in \bm{x}_i$ might be modeled as having a linear effect $z_i \cdot \beta$, a linear effect varying over time $z_i \cdot f(t)$, a smooth effect $f(z_i)$, or a smooth effect over time $f(z_i,t)$. For a detailed overview of different possible effect specifications including spatial or covariate interaction effects, see Table 2 of K\"ohler et al. (2017)\cite{kohler2017flexible}. We use suitable basis representations such as spline basis matrices with corresponding penalties to model the structured additive predictors as explained in detail in Section \ref{subsec:EstimationBasisRep}.
In particular, we further separate the longitudinal predictors 
\begin{align}
\label{eq:LongiPredictor}
    \eta_{\mu_k i}(t) = \sum_{h=1}^{H_{\mu_k} -1} f_{kh}(\bm{x}_i, t) + b_{i}^{(k)}(t)
\end{align}
into outcome specific (functional, i.e.\ potentially time-varying) fixed effects $f_{kh}(\bm{x}_i, t) \in L^2(\mathcal{I}), h = 1,...,H_{\mu_k} -1$, and subject and outcome specific functional random effects $b_{i}^{(k)}(t)$, which account for the correlation of longitudinal outcome values within a subject over time and across different markers.\cite{scheipl2015functional, volkmann2023multivariate} To model this correlation structure, we assume that the multivariate functional random effects $\bm{b}_i(t) = (b_i^{(1)}(t), ..., b_i^{(K)}(t))^{\top} \in L_K^2(\mathcal{I})$ with $L_K^2(\mathcal{I}) = L^2(\mathcal{I}) \times ... \times L^2(\mathcal{I})$ are independent copies of a smooth zero-mean Gaussian random process with multivariate covariance kernel $\mathcal{K}$. The associated covariance operator $\Gamma: L^2_K(\mathcal{I}) \rightarrow L^2_K(\mathcal{I})$ given by $(\Gamma \bm{g})(t) = \langle\langle\bm{g}, \mathcal{K}(t, \cdot)\rangle\rangle$ is based on the (potentially weighted) scalar product 
\begin{align}
\label{eq:scalarproduct}
\langle\langle \bm{g}, \bm{h} \rangle\rangle := \sum_{k=1}^K w_k \int_{\mathcal{I}}g^{(k)}(t)h^{(k)}(t)dt, \quad \bm{g},\bm{h} \in L_{K}^2(\mathcal{I})
\end{align} 
for given positive weights $w_k, k = 1,...,K$, inducing a Hilbert space structure\cite{happ2018multivariate}. Setting weights unequal to one allows to let variation from different longitudinal outcomes contribute differently to the multivariate kernel, which will be discussed in more detail in Section \ref{subsec:EstimationMFPCA}.
The corresponding covariance surfaces $\mathcal{K}^{(d, e)}(s, t) = \text{Cov}\left[b_i^{(d)}(s), b_i^{(e)}(t)\right], d, e = 1,.., K$ and $s,t \in \mathcal{I}$, are auto-covariances for $d = e$ and cross-covariances for $d \neq e$.

The auto-covariances describe the correlation structure of the random effects within a longitudinal outcome. The covariance implied by a subject specific constant shift and linear time trend such as in a parametric random intercept -- random slope mixed model would be contained as a special case. We do not parametrically restrict the auto-covariance structure, however, and it can thus also represent more flexible nonparametric forms such as in K\"ohler et al. (2017)\cite{kohler2017flexible}. The cross-covariances capture the correlation structure between the different longitudinal outcomes. As these are hard to realistically specify a priori, especially for nonparametric random effects, we allow the cross-covariances to nonparametrically vary  with the pair of time-points considered. Furthermore, we assume the $\bm{b}_i(t)$ to be independent of the Gaussian noise process \eqref{eq:longi_model_assumpt}, which allows to separate the error variance from the diagonal of the smooth covariance kernel $\mathcal{K}$.\cite{yao2005functional}

\subsection{Functional principal component representation of random effects}

K\"ohler et al. (2017)\cite{kohler2017flexible} achieve a flexible (auto-)covariance structure for one longitudinal marker by representing the functional random effects using B-spline basis functions with random effect weights. For multiple longitudinal outcomes, however, such an approach can become prohibitive as the number of parameters quickly increases: Using $D$ basis functions for each $b_i^{(k)}(t)$ results in $n\cdot K \cdot D$ random basis weights with a $(K \cdot D) \times (K \cdot D)$ unstructured covariance matrix $\bm{\Sigma}$ containing $(K \cdot D+1)K\cdot D / 2$ unknown parameters. Given that the desired flexibility is in practice often achieved by penalizing a large number $D$ of basis coefficients, a resulting \gls{mjm} suffers from the considerable number of parameters to estimate. On the other hand, reducing the number $D$ or assuming a diagonal covariance matrix, or both, are simplifying but unrealistic assumptions which could impair the estimation of the association with the event hazard.

Instead, we propose to use a functional principal components based representation of the multivariate functional random effects. We apply the multivariate \gls{kl}  theorem\cite{happ2018multivariate}, which allows to represent multivariate functional data in an infinite basis expansion using random scores and the eigenfunctions of the corresponding covariance operator. For the multivariate functional random effects we get \begin{align}
\label{eq:mfpc_represtation}
    \bm{b}_i(t) =  \sum_{m = 1}^{\infty}\rho_{im}\bm{\psi}_m(t) \approx \sum_{m = 1}^{M}\rho_{im}\bm{\psi}_m(t)
\end{align}
with random scores $\rho_{im}$ and the eigenfunctions $\bm{\psi}_m(t) = (\psi_m^{(1)}(t), ..., \psi_m^{(K)}(t))^{\top}\in L_K^2(\mathcal{I})$ of the covariance kernel $\mathcal{K}$\cite{happ2018multivariate, volkmann2023multivariate}, i.e.\ due to Mercer's theorem $\mathcal{K}(s,t) = \sum_{m=1}^\infty \nu_m \bm{\psi}_m(s) \bm{\psi}_m(t)^{\top}$ with ordered eigenvalues $\nu_1 \geq \nu_2  \geq \dots \geq 0$. Note that the eigenfunctions $\bm{\psi}_m, m = 1,..., \infty$ are orthonormal to each other with respect to the scalar product \eqref{eq:scalarproduct}. The scores $\rho_{im} \sim (0, \nu_m)$ are uncorrelated across $m$ and, in particular, for the Gaussian process $\bm{b}_i(t)$ are \gls{iid}  $\rho_{im} \sim N(0, \nu_m)$. In practice, the infinite expansion in \eqref{eq:mfpc_represtation} is truncated at a finite level using the leading $M$ eigenfunctions, where the truncation order $M$ controls the approximation accuracy.
This truncated multivariate \gls{kl} representation has several benefits: Interpretability, dimension reduction, and the simplification of the scalar random effects covariance matrix to a diagonal matrix.

The leading eigenfunctions correspond to the main modes of variation in the data due to decreasing corresponding eigenvalues and can be interpreted as outcome specific characteristic trajectories corresponding to correlated patterns that are simultaneously present across all longitudinal outcomes. The direction and strength of expression of these characteristic trajectories for each subject depend on their individual score $\rho_{im}$. The eigenvalues $\nu_m$ corresponding to the eigenfunctions allow to quantify their contributions to the overall variation in the data given that $E(\langle\langle 
\bm{b}_i,\bm{b}_i\rangle\rangle) =\sum_{k=1}^{K} w_k \int_{\mathcal{I}}\text{Var}(b_i^{(k)}(t))dt = \sum_{m = 1}^{\infty}\nu_m$. This can be used to determine the quality of the approximation in \eqref{eq:mfpc_represtation}, e.g.\ by chosing $M$ such that a pre-specified large percentage of variation is retained. We estimate the eigenfunctions as \glspl{mfpc} in a preliminary step described in section \ref{subsec:EstimationMFPCA}, leaving only the random scores $\rho_{im}$  to be estimated as random effects in the \gls{mjm}. This considerably reduces the number of parameters from $n\cdot K \cdot D$ to $n \cdot M$: The scores act as weights for multivariate functions and are thus not outcome-specific. In addition, $M$ only depends on the complexity of the covariance kernel $\mathcal{K}$, which is not necessarily connected to the number of longitudinal outcomes $K$, such that typically  in practice $M \ll K\cdot D$. Using the orthonormal \glspl{mfpc} as basis functions for the Gaussian random process also leads to \gls{iid} $\bm{\rho}_{i} = (\rho_{i1}, ..., \rho_{iM})^{\top} \sim N_M(\bm{0}, diag(\nu_1,..., \nu_M)), i = 1,.., n$, where the mean of the $M$-dimensional multivariate normal distribution is the vector of zeros $\bm{0}$ and the covariance matrix is diagonal with eigenvalues $\nu_m$ as elements. This represents a further simplification, reducing the covariance parameters to $M \ll (K \cdot D+1)K\cdot D / 2$  independent of $K$, and implies computational advantages compared to an unstructured covariance matrix $\bm{\Sigma}$ of a B-spline approach as described above.

The eigenfunction based representation of the random effects gives the best (in terms of integrated squared error \cite{ramsay2005}) approximation to the longitudinal outcomes $\bm{b}_i(t)$ for a given number of basis functions and allows a parsimonious model formulation. While in practice we have to use estimates of the basis functions $\bm{\psi}_m(t)$ in the \gls{mjm}, we are less interested in the \glspl{mfpc} themselves, although they provide additional interpretations of main directions of variation in marker trajectories. Our simulations show that even imperfectly estimated versions still serve as a good empirical basis for the longitudinal trajectories and give the \gls{mjm} the flexibility needed for a good model fit, yielding better results than more restrictive parametric or spline-based approaches in settings with nonlinear trends over time.

\section{Estimation}
\label{sec:Estimation}

We use a derivative-based \gls{mcmc} algorithm adapted from the one for univariate joint models proposed by K\"ohler et al. (2017)\cite{kohler2017flexible} to estimate the posterior distribution  of the parameter vector $\bm{\theta}$ containing all model parameters of interest  in a Bayesian framework. Note that the \gls{mfpc} basis is estimated directly from the data and is thus not part of $\bm{\theta}$. This might be interpreted as an empirical Bayes approach; however, we prefer to view the \glspl{mfpc} simply as a beneficial empirical basis for the random effects (comparable to e.g.\ pre-defined B-spline basis functions). The following sections outline the estimation of the \gls{mfpc} basis and the Bayesian estimation of $\bm{\theta}$.

\subsection{Estimation of the MFPC basis}
\label{subsec:EstimationMFPCA}

We estimate the \gls{mfpc} basis following the \gls{mfpca} proposed by Happ and Greven (2018)\cite{happ2018multivariate}, which estimates the multivariate covariance kernel $\mathcal{K}$ from separate \glspl{ufpca} based on the different longitudinal outcomes of each subject. This approach is very flexible as it allows to specify the \gls{ufpca} for each longitudinal outcome individually and relies only on the covariance between the estimated univariate scores of the subjects. 

In a first step, we estimate a mean trajectory $\eta_{\mu_k}^{*}(t_{ijk})$ for each longitudinal outcome using the outcome specific fixed effects $\sum_h^{H_{\mu_k} - 1} f_{kh}(\bm{x}_i, t)$ specified for the \gls{mjm} under a working independence assumption for the longitudinal observations ignoring censoring. We then estimate the auto-covariance of each longitudinal outcome $y_{ijk}(t) \in L^2(\mathcal{I})$ as a bivariate surface using fast symmetric additive covariance smoothing based on the crossproducts of the centered longitudinal observations $y_{ijk}^{*} = y_{ijk} - \eta_{\mu_k}^{*}(t_{ijk})$.\cite{cederbaum2018fast} A truncated eigendecomposition of the covariance surface then yields estimates for the first $M_k$ univariate eigenfunctions $\phi_{km_k} \in L^2(\mathcal{I})$ and eigenvalues corresponding to the variances of the scores $\xi_{km_ki}$ in the univariate \gls{kl} expansion $y_{ijk}(t) \approx \eta_{\mu_k}^{*}(t) + \sum_{m_k = 1}^{M_k}\xi_{km_ki}\phi_{km_k}(t)$. Given the estimated principal components $\hat\phi_{km_k}$, the univariate scores $\xi_{km_ki}$ are predicted using a conditional expectation approach.\cite{yao2005functional} Other approaches of estimating the \gls{ufpca} are also possible (cf.\ Section \ref{sec:Discussion}).

For a Hilbert space induced by \eqref{eq:scalarproduct}, Happ and Greven (2018)\cite{happ2018multivariate} show that the multivariate eigenfunctions are given as linear combinations of their univariate counterparts and derive the explicit formula for this relationship as $\psi^{(k)}_{m}(t) = (w_k)^{-1/2}\sum_{m_k = 1}^{M_k}c_{km_k}\phi_{km_k}(t)$, with linear combination weights $c_{km_k}$ depending on (the eigenvectors of) the covariance matrix of the univariate scores $\xi_{km_ki}, k = 1,...,K$.\cite{happ2018multivariate} The \gls{mfpca} thus uses the estimated covariance structure of the estimated univariate scores of all longitudinal outcomes to estimate the \glspl{mfpc} and consequently the multivariate covariance kernel $\mathcal{K}$. Regarding the choice of weights in \eqref{eq:scalarproduct}, setting $w_1 = ... = w_K = 1$ can be a reasonable starting point for the analysis in many cases. When the range or variation of the longitudinal data differ considerably between the longitudinal outcomes, we propose to use weights inversely proportional to the integrated univariate variance as given by the sum of univariate eigenvalues from the \glspl{ufpca}.\cite{happ2018multivariate} This approach is equivalent to standardization of the different longitudinal markers to make them comparable, as commonly also done in non-functional multivariate principal component analysis, while avoiding rescaling of the data. The resulting \glspl{mfpc} are then often more balanced across the longitudinal outcomes, avoiding \glspl{mfpc} which are dominated by one longitudinal outcome essentially capturing variation of that outcome alone. Refraining from the rescaling of longitudinal outcomes has the additional advantage that the association structure of the \gls{mjm} via the current value of the longitudinal outcomes can be interpreted much more intuitively.

As the \gls{mfpc} basis is an important part of the subsequent MJM, its estimation should be closely monitored in practice. It is important that the \glspl{ufpca} are flexible enough to capture all the relevant modes of variation in the data. In particular, the univariate truncation order $M_k$ should be high and a sufficient number of basis functions for the estimation of the univariate covariance surface should be specified, as variation lost by oversmoothing on the univariate level is lost for all  later stages of the analysis. In addition, the penalization chosen by the fast symmetric additive covariance smoothing directly influences the smoothness of the \glspl{mfpc} and consequently also the final fit of the \gls{mjm}. Visually inspecting the estimated \gls{mfpc} basis and consulting a subject matter expert about their beliefs on systematic variation in the data at this stage of the analysis can be helpful. 

Note that the estimation of the \gls{mfpc} basis uses the estimated covariance between the estimated univariate scores. As univariate scores are estimated with large uncertainty if only very few longitudinal observations are available, 
we recommend a trimmed approach where we use all longitudinal observations of subjects with at least one measurement after 10\% of the time interval $\mathcal{I}$ on all longitudinal outcomes. In our experience, this approach avoids extremely outlying estimates of univariate scores in the \gls{ufpca} for subjects with very early censoring, which via the estimated covariance between the univariate scores  can have a detrimental effect on the estimated \gls{mfpc} basis. Alternatively, outlying scores can be manually identified and  removed from the \gls{mfpca}, or robust alternatives to trimming such as weighting would also be possible.

Given a reasonable estimate of the \gls{mfpc} basis, the multivariate eigenvalues $\nu_m$ of the \gls{mfpca} can be used to choose the truncation order $M$ of the multivariate functional random effects representation \eqref{eq:mfpc_represtation}. The maximal number $M^{*} = \sum_{k=1}^{K}M_k$ of \glspl{mfpc} corresponds to the total number of \glspl{ufpc} in the \gls{mfpca}. Reducing $M < M^{*}$ allows further dimension reduction for the parameter vector $\bm{\theta}$ with relatively little loss in the estimation quality for high $M$ as shown in Section \ref{sec:Simulations}. A common criterion is to use a threshold for the ratio of cumulative explained variance, e.g.\ $99\%$, and discard all trailing \glspl{mfpc}. Given the \gls{mfpc} basis, choosing $M$ thus reduces to a model selection problem and can also be solved using standard approaches such as comparing information criteria for different model fits.

\subsection{Basis representation of effects}
\label{subsec:EstimationBasisRep}

The proposed \gls{mjm} is very flexible with respect to the readily available types of covariate effects for all structured additive predictors $\eta_{li}(t) = \sum_{h=1}^{H_l}f_{lh}(\bm{x}_i,t), l \in \mathcal{L}$. The Bayesian additive mixed models framework allows a wide range of effects such as linear, smooth, or random effects terms by using corresponding basis function expansions for $f_{lh}(\bm{x}_i,t)$ combined with associated priors.\cite{umlauf2018bamlss} All of these flexible effects, including the functional principal component based random effects, can be represented in matrix notation as
\begin{align}
\label{eq:basis_rep}
    \bm{f}_{lh} = \bm{X}_{lh}\bm{\beta}_{lh}
\end{align}
with vector $\bm{f}_{lh}$ of stacked function evaluations over all subjects for the survival submodel (vector length $n$) or over all subjects, markers and time points for the longitudinal submodel (vector length $N = \sum_i\sum_k n_{ik}$), design matrix $\bm{X}_{lh}$, and coefficient vector $\bm{\beta}_{lh}$. This representation includes but is not limited to  non-linear covariate effects, which can be represented as such a matrix product with  the columns of $\bm{X}_{lh}$ containing e.g.\ evaluations of spline basis functions. For the representation of further model terms, see Umlauf et al. (2018)\cite{umlauf2018bamlss} and K\"ohler et al. (2017)\cite{kohler2017flexible}. Note that standard sum-to-zero constraints are used to center the design matrix $\bm{X}_{lh}$ of non-linear functions\cite{scheipl2015functional, wood2017generalized} in order to ensure model identifiability in a model also including an intercept. 

In particular, the functional principal components based random effects $b^{(k)}_i(t)$ for one longitudinal outcome in \eqref{eq:LongiPredictor} can be expressed as
\begin{align}
\label{eq:re_blockdiag}
    \bm{b}^{(k)} = \bm{X}_{\mu_k H_{\mu_k}} \bm{\beta}_{\mu_k H_{\mu_k}}= blockdiag(\bm{\Psi}_1^{(k)},..., \bm{\Psi}_n^{(k)})\bm{\rho},
\end{align}
where the $N_k = \sum_i n_{ik}$-vector $\bm{b}^{(k)}$ contains the evaluations of the subject and outcome specific functional random effect at the observed time points for outcome $k$ and all objects $i$ and $\bm{\rho} = (\bm{\rho}_1^{\top},..., \bm{\rho}_n^{\top})^{\top}$ is the vector containing $M$ random scores $\bm{\rho}_i$ per subject (corresponding to the coefficient vector $\bm{\beta}_{\mu_k H_{\mu_k}}$). The design matrix $\bm{X}_{\mu_k H_{\mu_k}}$ has a blockdiagonal structure with blocks $\bm{\Psi}_i^{(k)}$ per $i$, which is an $n_{ik}\times M$ matrix containing evaluations of the $k$-th component of the $M$ \glspl{mfpc} at time points $t_{ijk}$, $j=1, \dots, n_{ik}$. Note that $\bm{\rho}$ is not specific to outcome $k$ and its dimension thus does not scale with $K$ due to our dimension reduction approach. Consequently, stacking the longitudinal outcomes to $\bm{b} = (\bm{b}^{(1)\top},..., \bm{b}^{(K)
\top})^{\top}$ simply corresponds to  stacking the $K$ blockdiagonal design matrices to a full multivariate longitudinal design matrix $\bm{X}_{\mu H_\mu} = (\bm{X}_{\mu_1 H_{\mu_1}}^{\top}, ..., \bm{X}_{\mu_K H_{\mu_K}}^{\top})^{\top}$ with a single coefficient vector $\bm{\rho}$. We point out that using a standard (spline) random effects basis for $\bm{b}$ would not be as parsimonious. While for each $k$ the design matrix $\bm{X}_{\mu_kH_{\mu_k}}$ would have a similar blockdiagonal structure as in \eqref{eq:re_blockdiag} with blocks containing the evaluations of the random effect (spline) basis, the random coefficient vector $\bm{\beta}_{\mu_k H_{\mu_k}}$ would be different for each $k$. As a result, combining the longitudinal outcomes would be equivalent to constructing a full multivariate longitudinal design matrix $\bm{X}_{\mu H_\mu} = blockdiag(\bm{X}_{\mu_1 H_{\mu_1}}, ..., \bm{X}_{\mu_K H_{\mu_K}})$.

\subsection{Priors}

In order to achieve the penalization of flexible effects in a Bayesian setting, appropriate priors for $\bm{\beta}_{lh}$ are specified. In general, we assume multivariate normal priors
\begin{align}
\label{eq:mvn_prior}
    p(\bm{\beta}_{lh} \mid \tau^2_{lh}) \propto (\tau_{lh}^2)^{-\frac{\text{rank}(\bm{K}_{lh})}{2}}\exp\left(-\frac{1}{2\tau^2_{lh}}\bm{\beta}_{lh}^{\top} \bm{K}_{lh}\bm{\beta}_{lh}\right)
\end{align}
with variance parameter $\tau_{lh}^2$, $\text{rank}(\bm{A})$ denoting the rank of matrix $\bm{A}$, and $\bm{K}_{lh}$ the chosen precision matrix. The variance parameter takes the role of an inverse smoothing parameter controlling the trade-off between flexibility and smoothness of the flexible effect, while the precision matrix determines the type of penalization and contains e.g.\ difference penalties for Bayesian P-splines.\cite{umlauf2018bamlss} We assume independent inverse Gamma hyperpriors for the variance parameters $\tau_{lh}^2 \sim IG(0.001, 0.001)$, which results in inverse Gamma full conditionals but other priors, such as half-Cauchy, are also possible. Furthermore, priors for anisotropic smooths containing multiple variance parameters can be specified as given in K\"ohler et al. (2017) \cite{kohler2017flexible}. For linear or parametric time-constant terms, we approximate $\bm{K}_{lh} = \bm{0}$ by using $\bm{\beta}_{lh} \sim N(\bm{0}, 1000^2\bm{I})$ with identity matrix $\bm{I}$.

Following the truncated \gls{kl}  representation \eqref{eq:mfpc_represtation}, we can assume independent random scores over the subjects $i$ and the \glspl{mfpc} $m$ so that we can separate $\bm{\rho} = (\bm{\rho}_{(1)},..., \bm{\rho}_{(M)})^{\top}$ with $\bm{\rho}_{(m)} = (\rho_{1m}, ..., \rho_{nm})^{\top}$ and use separate priors for each $\bm{\rho}_{(m)}$. In particular, we assume the multivariate normal priors \eqref{eq:mvn_prior} for each $m$, with precision matrix $\bm{K}_{(m)} = \bm{I}$ and different variance parameters $\tau_{(m)}^2$. Note that the variances $\tau_{(m)}^2$ correspond to the eigenvalues $\nu_m$ in the \gls{mfpca}. We emphasize viewing the estimated \glspl{mfpc} as an empirical basis by using the above introduced weak Gamma priors for the $\tau_{(m)}^2$ instead of using the estimated eigenvalues $\nu_m$ in the priors. With our specification, the $\rho_{im}$ take the role of random effects for the basis functions $\bm{\psi}_m$ - defined simultaneously  for all $k$ on the entire follow-up time interval $\mathcal{I}$ - as a flexible data-driven basis, replacing parametric basis functions such as e.g.\ a constant and a linear function per $k$ in a random intercept -- random slope model.

\subsection{Likelihood}

We use the standard assumption of shared random effects models that given the parameter vector $\bm{\theta}$, the longitudinal outcomes $\bm{y} = (\bm{y}_1^{\top}, ..., \bm{y}_K^{\top})^{\top}$  and the survival outcome $[\bm{T}, \bm{\delta}]$ are conditionally independent. 
Here, $\bm{y}_k = (\bm{y}_{1k}^{\top}, ..., \bm{y}_{nk}^{\top})^{\top}, \bm{y}_{ik} = (y_{i1k},..., y_{in_{ik}k})^{\top}$,  $\bm{T} = (T_1,..., T_n)^{\top}$ the follow-up (event or censoring) times, and the event indicators $\bm{\delta} = (\delta_1,...,\delta_n)^{\top}$ are 1 if the subject experiences the event and 0 for censoring. 
This allows to factorize the likelihood of the \gls{mjm} as the product of the Likelihood $L^{surv}$ of survival submodel \eqref{eq:hazard_one_observation} and the Likelihood $L^{long}$ of the longitudinal submodel \eqref{eq:longi_model_one_obs}-\eqref{eq:longi_model_assumpt}. The log-likelihood of the \gls{mjm} can then be formulated as
\begin{align}
    \ell(\bm{\theta} \mid \bm{T}, \bm{\delta}, \bm{y}) = \ell^{surv}(\bm{\theta} \mid \bm{T}, \bm{\delta}) + \ell^{long}(\bm{\theta} \mid \bm{y})
\end{align}
with the log-likelihoods $\ell^{surv}$ and $\ell^{long}$ of the survival and longitudinal submodels, respectively.

For the survival submodel, the log-likelihood can be expressed as
\begin{align*}
    \ell^{surv}(\bm{\theta} \mid \bm{T}, \bm{\delta}) = \bm{\delta}^{\top}\bm{\eta}(\bm{T}) - \bm{1}_n^{\top}\bm{\Lambda}(\bm{T}),
\end{align*}
where $\bm{\eta}(\bm{T}) = (\eta_1(T_1), ..., \eta_n(T_n))^{\top}$, $\bm{1}_u$ is the $u$-vector of ones, and $\bm{\Lambda}(\bm{T}) = (\Lambda_1(T_1),..., \Lambda_n(T_n))^{\top}$ is the vector of cumulative hazard rates $\Lambda_i(T_i) = \int_0^{T_i}\eta_i(s)ds$.

The log-likelihood of the longitudinal submodel is similar to a multivariate functional additive mixed model \cite{volkmann2023multivariate} with the longitudinal outcomes stacked in the outcome vector $\bm{y}$ and
\begin{align*}
\ell^{long}(\bm{\theta} \mid \bm{y}) & = -\frac{N}{2}\log(2\pi) - \bm{1}_{N}^{\top}\bm{\eta}_{\sigma} - \frac{1}{2}(\bm{y}- \bm{\eta}_{\mu})^{\top}\bm{R}^{-1}(\bm{y}- \bm{\eta}_{\mu})
\end{align*}
with similarly stacked vectors $\bm{\eta}_{\sigma} = (\bm{\eta}_{\sigma_1}^{\top},..., \bm{\eta}_{\sigma_K}^{\top})^{\top}$, $\bm{\eta}_{\sigma_k}=(\bm{\eta}_{\sigma_k 1}^{\top},..., \bm{\eta}_{\sigma_k n}^{\top})^{\top}$, $\bm{\eta}_{\sigma_k i}=(\eta_{\sigma_ki}(t_{i1k}),..., \eta_{\sigma_ki}(t_{in_{ik}k}))^{\top}$ and $\bm{\eta}_{\mu} = (\bm{\eta}_{\mu_1}^{\top},..., \bm{\eta}_{\mu_K}^{\top})^{\top}$, $\bm{\eta}_{\mu_k}=(\bm{\eta}_{\mu_k1}^{\top},..., \bm{\eta}_{\mu_kn}^{\top})^{\top}$,$ \bm{\eta}_{\mu_ki} = (\eta_{\mu_ki}(t_{i1k}),...,\eta_{\mu_ki}(t_{in_{ik}k}))^{\top}$, and diagonal matrix $\bm{R} = diag(\exp(\bm{\eta}_{\sigma})^2)$.

\subsection{Posterior mean estimation}

The posterior of the full parameter vector $\bm{\theta}$ factorizes to
\begin{align*}
p(\bm{\theta}\mid \mathbf{T}, \bm{\delta}, \mathbf{y}) & \propto L^{surv}(\bm{\theta} \mid \mathbf{T}, \bm{\delta}) \cdot
L^{long}(\bm{\theta} \mid \bm{y})  \cdot \prod_{l\in \mathcal{L}} \prod_{h=1}^{\tilde{H}_l} (p(\bm{\beta}_{lh}\mid \bm{\tau}_{lh}^2)p(\bm{\tau}_{lh}^2))\cdot \prod_{m = 1}^{M}(p(\bm{\rho}_{(m)}\mid\bm{\tau}_{(m)}^2)p(\bm{\tau}_{(m)}^2)),
\end{align*}
where $\tilde{H}_l = H_l - 1$ for the longitudinal outcomes $l \in \{\mu_1,..., \mu_K\}$ and $\tilde{H}_l = H_l$ otherwise. We obtain starting values for the derivative-based \gls{mcmc} algorithm by approximating the mode of the posterior using a blockwise Newton-Raphson procedure.\cite{kohler2017flexible} In contrast to K\"ohler et al. (2017), however, we  use the approximate mode only as a starting value for mean estimation, thus prioritizing the fast convergence of the Newton-Raphson algorithm. The step length is set to 1 except for the survival predictors $l \in \{\lambda, \gamma\}$, where it is set to 0.1, and the variance parameters are initialized by the \textbf{bamlss} algorithm so that the empirical degrees of freedom are comparable across model terms.\cite{umlauf2021bamlss,pkg:bamlss} 

Then, the \gls{mcmc} algorithm uses the blockwise score vectors $\bm{s}(\bm{\beta}_{lh})$ and Hessian $\bm{\mathcal{H}}(\bm{\beta}_{lh})$ of the log-posterior given in Appendix \ref{app:derivatives} to approximate the full conditionals $p(\bm{\beta}_{lh}\mid \cdot)$.\cite{umlauf2018bamlss} These approximations are based on a second-order Taylor expansion of the log-posterior centered at the last state $\bm{\beta}_{lh}^{[u]}$ for \gls{mcmc} iteration $u$ and result in a proposal density proportional to a multivariate normal distribution $N(\bm{\mu}_{lh}^{[u]}, \bm{\Sigma}_{lh}^{[u]})$ with mean $\bm{\mu}_{lh}^{[u]} = \bm{\beta}_{lh}^{[u]} - \bm{\mathcal{H}}(\bm{\beta}_{lh}^{[u]})^{-1}\bm{s}(\bm{\beta}_{lh}^{[u]})$ and precision matrix $(\bm{\Sigma}_{lh}^{[u]})^{-1} = -\bm{\mathcal{H}}(\bm{\beta}_{lh}^{[u]})$. While the calculation of the derivatives increases computational cost, it has the advantage that this \gls{mcmc} algorithm approximates a Gibbs sampler, resulting in high acceptance rates and good mixing. Gibbs sampling is used for variance parameters $\tau_{lh}^2$ with conjugate inverse Gamma hyperpriors and slice sampling otherwise.

\subsection{Implementation}

The implementation of the proposed approach builds on the flexibility provided by the R package \textbf{bamlss}\cite{umlauf2021bamlss,pkg:bamlss}. We supply the package \textbf{MJMbamlss} on Github (\url{https://github.com/alexvolkmann/MJMbamlss})
containing a new family-class for the presented \gls{mjm} as well as some convenience functions for estimating the \gls{mfpc} basis. The contained vignettes illustrate how to combine the two packages for model estimation. Similar to the package \textbf{JMbayes2}\cite{rizopoulos2022jmbayes2} we find that standardizing the survival design matrices within the estimation process improves mixing and convergence of the \gls{mcmc} algorithm as well as posterior mode estimation, and is implemented by default. The integrals in the cumulative hazard rates $\Lambda_i(T_i)$ and in the corresponding derivatives are of non-standard form and we use seven point Gaussian quadrature to numerically approximate them.\cite{press2007numeric}

\section{Simulations}
\label{sec:Simulations}

We show the good performance of our proposed approach in two simulation scenarios. Scenario I aims at comparing our approach to the \textbf{JMbayes2}\cite{rizopoulos2022jmbayes2} package in a parametric setting that is favorable to \textbf{JMbayes2}, i.e.\ in a setting with a simple random intercept -- random slope covariance structure, here for six longitudinal outcomes. Our approach, estimating the random effects structure nonparametrically, should be able to also recover this truly linear structure, even though a higher variance compared to correctly specifying the parametric model in \textbf{JMbayes2} is to be expected due to the increased model flexibility. This example also illustrates the reduction of parameters to estimate in our \gls{mfpc} based representation, while keeping a comparable estimation performance. Scenario II by contrast focuses on a complex covariance structure, here between two nonlinear longitudinal markers, and highlights the powerful and parsimonious modeling possibilities of \glspl{mfpc} for highly flexible longitudinal trajectories.
In the following, the data generating processes of the two simulation scenarios are described, the general setup of the simulation is introduced, and the results are presented. Full details of the data generation, such as the used parameter values, as well as additional results are provided in Appendix \ref{app:simulation}.

\subsection{Scenario I}

We generate a sequence of six different longitudinal trajectories for $n = 150$ subjects at an equidistant grid of 101 time points $\mathcal{P} \in [0,1]$. The longitudinal trajectories follow a linear predictor with a fixed linear time trend and linear effect of a binary covariate $x_{gi}$, sampled from a discrete uniform distribution, as well as a random intercept and random slope over time. The fixed effects are chosen identical over the different longitudinal outcomes, but the random effects differ considerably in magnitude. We draw the random effects from a 12-dimensional multivariate normal distribution with mean zero and given unstructured covariance matrix. While the correlation between random intercept and slope within a longitudinal outcome is the same for all outcomes, the scale of the variances and thus covariances increases with $k$. At the same time, the correlation between outcomes $k$ and $l$ is specified as a decreasing function of their difference, i.e.\ ``neighbouring'' outcomes have a higher correlation, and equal to zero for the furthest pairs (12 of the 78 covariance parameters are zero, corresponding to independence of random effects). We specify the linear predictors $\eta_{\gamma}$ as containing the binary covariate $x_{gi}$, the baseline hazard $\eta_{\lambda}$ as following a Weibull distribution, and $\eta_{\alpha_1},...,\eta_{\alpha_6}$ as constants with decreasing values from $1.5$ to $-1.5$. The survival times are then generated by calculating the hazard rate and deriving the corresponding survival probabilities.\cite{bender2005generating, crowther2013simulating} Random censoring is introduced by drawing censoring times from a uniform distribution on $[0, 1.75]$. The subject specific follow-up time $T_i$ is the minimum of the survival time, the censoring time and 1. We introduce subject and outcome specific observation times by sampling $25\%$ of the remaining subject specific subset of $\mathcal{P}$ with a maximum of $15$ longitudinal observations per outcome. The longitudinal observations are then generated by adding the Gaussian noise $\epsilon_{ijk}$ with variance constant across $k$ to the linear predictor evaluated at the observation times. In the generated data sets, all subjects have baseline measurements of the longitudinal outcomes at time point 0 so that the subjects have a minimum of six (one per outcome) and up to $90$ (mean $63$) longitudinal observations across all outcomes. The data sets show a mean event rate of $43\%$, with mean follow-up time of $0.52$.

\subsection{Scenario II}

For simulation scenario II, we follow the same data generation structure to obtain data sets with $n = 300$ subjects and two longitudinal outcomes based on different coefficient specifications for the additive predictors compared to scenario I and a random censoring uniform distribution on $[0, 3]$. Most notably, the random intercept and slope components in the longitudinal additive predictors are replaced by a more complex covariance structure. We include multivariate functional random effects drawn from a multivariate Gaussian process with a covariance kernel based on six multivariate eigenfunctions. To do this, we follow the results for finite \gls{kl} decompositions of Happ and Greven (2018)\cite{happ2018multivariate}, and define univariate non-orthogonal basis functions and a covariance matrix for the coefficients of the corresponding basis expansion to construct the multivariate covariance kernel. We specify three basis functions per longitudinal outcome as splines with distinct characteristics at the beginning, middle, or end of the interval $[0,1]$: For outcome $1$ the basis functions correspond to level shifts, whereas for outcome $2$ the basis functions describe short-term peaks (see Appendix \ref{app:simulation} Figure \ref{APPfig:sim_basis_fcts}). The $6\times 6$ covariance matrix of the basis coefficients is generated from an eigenvalue decomposition with specified eigenvalues and a random orthonormal matrix. This gives a non-standard multivariate covariance kernel with highly nonlinear eigenfunctions (see Appendix \ref{app:simulation} Figure \ref{APPfig:sim_eigenfcts}). To generate multivariate functional random effects from this Gaussian process, we draw \gls{iid} samples $\bm{\rho}_i = (\rho_{i1},...,\rho_{i6})^{\top} \sim N_6(\bm{0}, diag(\nu_1,...,\nu_6)), i = 1,...,300$ with corresponding multivariate eigenvalues $\nu_1,...,\nu_6$. The data sets in scenario II have on average 24 longitudinal observations (minimum 2 and maximum 30) and a mean event rate of $57\%$ with mean follow-up time 0.6.

\subsection{Simulation Setup}

For both scenarios we simulate 200 data sets to which we apply four different model estimation approaches. First, we estimate a model using the true underlying covariance kernel via the corresponding true \gls{mfpc} basis in the \textbf{bamlss} framework to have an oracle benchmark model (denoted as \textit{TRUE}). Comparing the benchmark to a model where the covariance structure is estimated from the data as described in Section \ref{subsec:EstimationMFPCA} (\textit{EST}) allows us to quantify the loss in estimation accuracy due to the \gls{mfpc} basis estimation. We also visualize the impact of further reducing the number of included \gls{mfpc} based random effects by lowering the truncation order $M$ to at least $99\%$ explained variance of the Gaussian process in a third modeling approach (\textit{TRUNC}). Finally, we compare our proposed models to an appropriate model fit (depending on the scenario) in the widely used \textbf{JMbayes2} package (denoted as \textit{JMB}), which also uses Bayesian model estimation with priors chosen based on preliminary separate univariate model fits for the longitudinal and survival outcomes.\cite{rizopoulos2022jmbayes2} In particular, we incorporate the knowledge of truly linear random effects and specify a random intercept -- random slope model with unstructured covariance matrix for the \textit{JMB} model in scenario I. In scenario II, the covariance structure is modeled by including a subject and outcome specific random intercept as well as three cubic B-spline basis functions resulting in an $8\times 8$ unstructured covariance matrix. We acknowledge that this modeling approach might not be flexible enough for the specified covariance structure of the data generating process. However, increasing the number of basis functions to four already leads to numerical problems in more than $75\%$ of the \textbf{JMbayes2} simulation runs, whereas with the proposed number only one model cannot be estimated in \textbf{JMbayes2}. Note that the \textbf{bamlss} models do not show any numerical problems in our simulation. In both frameworks, the baseline hazard is modeled using cubic P-splines with third order difference penalty and 20 basis functions (before centering).

We run the \gls{mcmc} algorithm for each model for 5500 iterations and use a burnin of 500 samples and a thinnig of 5 to obtain a total of 1000 samples per model. The model fits are evaluated using bias, \gls{mse}, and frequentist coverage of the $95\%$ credible intervals of the additive predictors. In particular, we define the average bias for the longitudinal predictors $l \in \{\mu_k, \sigma_k| k = 1,...,K\}$ in simulation run $u = 1,..., 200$ as $Bias^{u}_{l} = \frac{1}{N_{k}^u}\sum_{i=1}^{n}\sum_{j = 1}^{n^u_{ik}} (\hat{\eta}^u_{li}(t_{ijk}^u) - \eta^u_{li}(t_{ijk}^{u}))$, with estimate $\hat{\eta}^u_{li}$ and the superscript $u$ denoting simulation run specific analogues. We also evaluate the model fit over time using a time dependent bias defined as $Bias^{u}_{l}(t) = \frac{1}{n_{risk}^u(t)}\sum_{i\in \mathcal{R}^u(t)} (\hat{\eta}^u_{li}(t) - \eta^u_{li}(t))$ for all $t$ in $\mathcal{P}$, where $n_{risk}^u(t)$ is the cardinality of the risk set of subjects $\mathcal{R}^u(t)$ at time $t$. Note that for the survival submodel, we evaluate the survival predictors $\eta_{\lambda}$ and $\eta_\gamma$ jointly using their sum $\eta_{\lambda + \gamma}$, as the two frameworks handle intercepts and sum-to-zero constraints for smooth effects differently. For $l \in \{\lambda + \gamma, \alpha_1, ..., \alpha_K\}$ we calculate the average bias $Bias^{u}_{l} = \frac{1}{n}\sum_{i=1}^{n}(\eta_{li}^u(T_i^{u}) - \hat{\eta}^u_{li}(T_i^u))$ at the subject specific follow-up times as well as the time dependent bias $Bias^{u}_{l}(t)$ as above. We define the \gls{mse} and the coverage analogously for all predictors, for instance $rMSE^{u}_{l} = (\frac{1}{n}\sum_{i=1}^{n}(\hat{\eta}^u_{li}(T_i^u) 
- \eta^u_{li}(T_i^u))^2)^{1/2}$ and $Coverage_l^u = \frac{1}{n}\sum_{i=1}^{n}I(\hat{\eta}_{li}^{u,2.5}(T_i^u) \leq \eta^u_{li}(T_i^u) \leq \hat{\eta}_{li}^{u,97.5}(T_i^u))$ with indicator function $I(\cdot)$ and $\hat{\eta}^{u,\alpha}$ the $\alpha\%$ quantile of the \gls{mcmc} samples for the posterior mean for the survival predictors $l \in \{\lambda + \gamma, \alpha_1, ..., \alpha_K\}$. Naturally, these definitions simplify correspondingly for the predictors $l \in \{\alpha_1,...,\alpha_K, \sigma_1, ..., \sigma_K\}$, which are constant over all subjects and time in our simulation.
The evaluation criteria of the 200 (or 199 for the \textbf{JMbayes2} approach in scenario II) simulation runs are then averaged for presentation.

Additionally, we evaluate the estimated \glspl{mfpc} in both scenarios. Eigenfunctions are generally only defined up to a sign change, which is why we first use the norm $||\cdot||$ induced by \eqref{eq:scalarproduct} (with weights chosen as $1$ throughout the simulation) to determine whether a reflected version $(-\hat{\bm{\psi}}_{m}(t))$ is closer to the true eigenfunction $\bm{\psi}_{m}(t), m = 1,...,M$. We then use the squared norm of the error, $||\bm{\psi}_{m} - \hat{\bm{\psi}}_{m}||^2$, to evaluate the estimated \gls{mfpc}.

\subsection{Simulation results}

Table \ref{tab:simI_results} contains the results of simulation scenario I. Comparing the \textbf{bamlss} model \textit{TRUE} with true underlying eigenfunctions as basis functions to the \textbf{JMbayes2} model \textit{JMB} allows to directly compare the two frameworks for \glspl{mjm}, contrasting the different choices of basis and correlation structure without influence of the particular method used to estimate the \glspl{mfpc}. More precisely, while both approaches predict the two random effects for each subject in their model fit, the \gls{mfpc} based approach only estimates the random effect variances whereas \textbf{JMbayes2} also estimates the correlation between the random intercepts and slopes on all longitudinal outcomes. Overall, we find that the two approaches yield similar bias and \gls{mse} values in the linear scenario I. In more detail, \textbf{bamlss} performs slightly better for the survival predictors $\eta_{\lambda} + \eta_{\gamma}$ and the longitudinal predictors $\eta_{\mu_1},...\eta_{\mu_6}$, while the \textbf{JMbayes2} approach results in better estimates of the association predictors $\eta_{\alpha_1},..., \eta_{\alpha_6}$. Aside from the different prior specifications, some of this difference in performance for the $\eta_{\alpha_k}$ might also be caused by the relatively short \gls{mcmc} chains to lower the computational burden of the simulation. The frequentist coverage of both frameworks is close to the nominal value of $95\%$ for all predictors except the combined survival predictor containing the baseline hazard. Bender et al.\ (2018)\cite{bender2018generalized} point out that using P-splines with a single penalty/variance parameter can be suboptimal when the baseline hazard changes quickly over the follow-up period and suggest to use adaptive spline smooths instead, which is only available (at increased computational cost) for \textbf{bamlss} and not for \textbf{JMbayes2}, and thus not used in this comparison.

\begin{table}
\caption{Results of simulation scenario I. The bias and \gls{mse} values for $\eta_{\mu_1},...,\eta_{\mu_6}$ are multiplied by factor $100$ for better readability. The values for $\eta_{\sigma_1},...,\eta_{\sigma_6}$ are averaged and reported as $\eta_\sigma$.}
\label{tab:simI_results}
\vspace{1em}
\centering
\footnotesize
\begin{tabular}{c|rrrr|rrrr|rrrr}
\hline
 & \multicolumn{4}{c|}{\textbf{Bias}} &\multicolumn{4}{c|}{\textbf{rMSE}} &\multicolumn{4}{c}{\textbf{Coverage}} \\
 $\eta$& \textit{TRUE} & \textit{EST} & \textit{TRUNC} & \textit{JMB} & \textit{TRUE} & \textit{EST} & \textit{TRUNC} & \textit{JMB}& \textit{TRUE} & \textit{EST} & \textit{TRUNC} & \textit{JMB}\\
\hline
$\lambda + \gamma$ & 0.811 & 0.812 & 0.812 & 1.006 & 0.907 & 0.911 & 0.909 & 1.124 & 0.438 & 0.433 & 0.436 & 0.465 \\ 
  $\alpha_{1}$ & 0.118 & 0.106 & 0.126 & 0.042 & 0.313 & 0.315 & 0.338 & 0.283 & 0.945 & 0.930 & 0.920 & 0.955 \\ 
  $\alpha_{2}$ & 0.014 & 0.014 & 0.012 & 0.003 & 0.256 & 0.261 & 0.267 & 0.242 & 0.950 & 0.955 & 0.965 & 0.970 \\ 
  $\alpha_{3}$ & 0.047 & 0.035 & 0.029 & 0.006 & 0.214 & 0.211 & 0.220 & 0.199 & 0.955 & 0.970 & 0.965 & 0.965 \\ 
  $\alpha_{4}$ &- 0.053 & -0.046 & -0.052 & -0.046 & 0.208 & 0.213 & 0.216 & 0.198 & 0.945 & 0.945 & 0.935 & 0.955 \\ 
  $\alpha_{5}$ & -0.061 & -0.056 & -0.054 & -0.051 & 0.201 & 0.199 & 0.199 & 0.186 & 0.965 & 0.955 & 0.955 & 0.970 \\ 
  $\alpha_{6}$ & -0.168 & -0.157 & -0.158 & -0.106 & 0.259 & 0.251 & 0.256 & 0.227 & 0.905 & 0.895 & 0.905 & 0.960 \\ 
  $\mu_{1}$ & 0.003 & 0.004 & 0.003 & 0.004 & 2.510 & 4.785 & 9.786 & 2.516 & 0.949 & 0.843 & 0.673 & 0.948 \\ 
  $\mu_{2}$ & 0.001 & 0.000 & -0.000 & -0.004 & 2.516 & 4.872 & 9.005 & 2.522 & 0.948 & 0.842 & 0.715 & 0.946 \\ 
  $\mu_{3}$ & -0.004 & -0.003 & -0.003 & -0.010 & 2.507 & 5.079 & 8.250 & 2.511 & 0.950 & 0.839 & 0.755 & 0.947 \\ 
  $\mu_{4}$ & 0.004 & 0.004 & 0.004 & 0.002 & 2.523 & 5.363 & 6.944 & 2.528 & 0.949 & 0.834 & 0.796 & 0.947 \\ 
  $\mu_{5}$ & -0.002 & -0.002 & -0.002 & -0.003 & 2.518 & 5.463 & 6.377 & 2.522 & 0.950 & 0.831 & 0.807 & 0.949 \\ 
  $\mu_{6}$ & 0.011 & 0.012 & 0.011 & 0.011 & 2.527 & 5.914 & 6.240 & 2.531 & 0.950 & 0.827 & 0.817 & 0.948 \\ 
  $\sigma$ & 0.002 & 0.263 & 0.467 & 0.001 & 0.016 & 0.263 & 0.467 & 0.016 & 0.945 & 0.056 & 0.028 & 0.938 \\ 
\hline
\end{tabular}
\end{table}

Estimating the \gls{mfpc} basis functions from the data results in a clear increase in the variance of the estimated longitudinal submodel, due to the higher flexibility of our nonparametric approach, which gives an advantage to the correctly specified parametric \textit{JMB} and \textit{TRUE} models (the latter using the true linear \glspl{mfpc}). The higher variability is reflected in higher \gls{mse} values of the \textit{EST} models, while the bias values for the longitudinal predictors remain stable. The reason is that flexibly estimating the eigenfunctions from the data in scenario I does not necessarily result in linear estimated \glspl{mfpc}  (see Appendix \ref{app:simulation} Figure \ref{APPfig:sim_scenI_mfpcs}), which can lead to longitudinal model fits that are too wiggly compared to the true parametric random intercept -- random slope data generating process assumed known in \textit{JMB} and \textit{TRUE} models. This mixing of the correlation structure and the random noise also leads to an overestimation of the variance predictors $\eta_{\sigma_1},..., \eta_{\sigma_6}$ (averaged in Table \ref{tab:simI_results}) coupled with poor frequentist coverage in the longitudinal submodel. This, however, does not negatively affect the estimation of the survival submodel where the bias, \gls{mse}, and coverage values are similar to the \textit{TRUE} models, including for the association predictors $\eta_{\alpha_1},...,\eta_{\alpha_6}$. Note that the bias and \gls{mse} values for the longitudinal predictors in Table \ref{tab:simI_results} have been multiplied by a factor $100$, such that the longitudinal model fits using the estimated \glspl{mfpc} still are visually  close  to the true trajectories (see Appendix \ref{app:simulation} Figure \ref{APPfig:sim_scenI_fits}) and seem close enough to correctly estimate the association predictors well. 

Lowering the number of \gls{mfpc} basis functions included in the model by introducing truncation at a prespecified rate of explained variance of the Gaussian process is expected to lower the computational burden, but to deteriorate the longitudinal submodel fit. In particular, for a truncation at $99\%$ explained variance, the \textit{TRUNC} models offer a notable dimension reduction, as most models only include ten \glspl{mfpc} (minimum nine and maximum eleven) instead of the full 12, thus cutting down the models by $2 \times 150$ parameters. While the bias values of the longitudinal predictors are basically unaffected by the truncation, we note that the \gls{mse} values of the longitudinal predictors are up to four times higher while the frequentist coverage drops to about $70 - 80\%$. The variation of the Gaussian process now unaccounted for by the excluded \glspl{mfpc} is absorbed by the variance predictors $\eta_{\sigma_1}, ..., \eta_{\sigma_6}$, which are overestimated throughout all \textit{TRUNC} models. The \textit{TRUNC} models show, however, that the trailing \glspl{mfpc} with small variances mainly affect  the fit of the longitudinal submodel (see Appendix \ref{app:simulation} Figure \ref{APPfig:sim_scenI_fits}) while the association predictors are still estimated reasonably well. Truncation thus allows an explicit choice and trade-off between computational complexity and (longitudinal) model fit, with the ordering of eigenvalues ensuring that components of little importance are truncated first.

\begin{figure}
    \centering
    \includegraphics[width=0.8\textwidth]{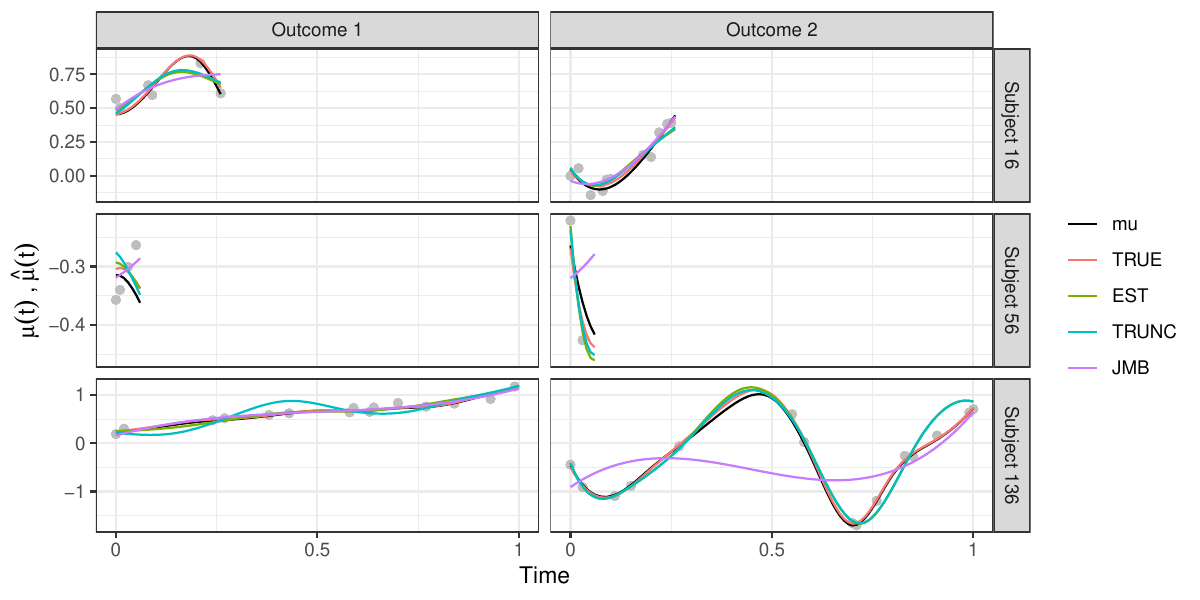}
    \caption{Different model fits of the longitudinal submodel for three random subjects of a simulated data set in scenario II. The solid black line corresponds to the true underlying longitudinal trajectory (mu) and the grey dots are the observed data points. The \textit{TRUNC} model in this example includes five \glspl{mfpc} instead of the correct six.}
    \label{fig:simII_modelfits}
\end{figure}

In simulation scenario II, the parametric covariance specification of scenario I is replaced by a complex multivariate covariance structure. In this setting,  the \textit{JMB} models based on the \textbf{JMbayes2} framework no longer have the advantage of ``knowing'' the correct parametric model. In fact, we find that the restricted flexibility of the \textit{JMB} models (as mentioned above) leads to a worse overall model fit compared to \textit{TRUE} models as measured by the bias, \gls{mse}, and frequentist coverage values for (almost) all predictors as given in Table \ref{tab:simII_results}. The difference between the two models is especially pronounced for the longitudinal predictors $\eta_{\mu_1}, \eta_{\mu_2}$, and $\eta_{\sigma_1}, \eta_{\sigma_2}$, but also for the association predictors $\eta_{\alpha_1}, \eta_{\alpha_2}$. A negative bias for positive values of the underlying association in simulation II indicates that the \textit{JMB} models underestimate the association parameters. Figure \ref{fig:simII_modelfits} illustrates for three random subjects that the \textit{TRUE} model using the underlying eigenfunctions as basis functions fits closely to the true trajectories, while the \textit{JMB} model tends to oversmooth the data, which may explain its slight underestimation of the $\alpha_1, \alpha_2$ association parameters also known to occur in regression settings with covariate measurement error\cite{carroll2006measurement}. Especially for subjects with a short follow-up time (such as subject $56$), principal component based random effects may be more capable of pooling the information over all subjects and across markers to approximate the true trajectories, while in the example of subject $56$ in Figure \ref{fig:simII_modelfits} the \textit{JMB} model even fits an incorrectly increasing time trend based on the few data points available.

\begin{table}
\caption{Results of simulation scenario II. The bias and \gls{mse} values for $\eta_{\mu_1},\eta_{\mu_2}$ are multiplied by factor $100$ for better readability.}
\label{tab:simII_results}
\vspace{1em}
\footnotesize
\centering
\begin{tabular}{c|rrrr|rrrr|rrrr}
\hline
 & \multicolumn{4}{c|}{\textbf{Bias}} &\multicolumn{4}{c|}{\textbf{rMSE}} &\multicolumn{4}{c}{\textbf{Coverage}} \\
 $\eta$& \textit{TRUE} & \textit{EST} & \textit{TRUNC} & \textit{JMB} & \textit{TRUE} & \textit{EST} & \textit{TRUNC} & \textit{JMB}& \textit{TRUE} & \textit{EST} & \textit{TRUNC} & \textit{JMB}\\
\hline
$\lambda + \gamma$ & 1.087 & 1.129 & 1.135 & 1.473 & 1.185 & 1.223 & 1.228 & 1.545 & 0.154 & 0.147 & 0.148 & 0.103 \\ 
  $\alpha_{1}$ & 0.019 & 0.012 & 0.012 & -0.053 & 0.070 & 0.072 & 0.073 & 0.098 & 0.945 & 0.935 & 0.950 & 0.849 \\ 
  $\alpha_{2}$ & 0.010 & 0.005 & 0.002 & -0.043 & 0.070 & 0.074 & 0.079 & 0.104 & 0.990 & 0.960 & 0.955 & 0.859 \\ 
   $\mu_{1}$ & -0.004 & -0.004 & -0.006 & -0.002 & 2.894 & 11.271 & 12.808 & 13.013 & 0.944 & 0.795 & 0.770 & 0.810 \\ 
  $\mu_{2}$ & -0.027 & -0.028 & -0.027 & -0.036 & 2.850 & 8.917 & 9.752 & 20.246 & 0.934 & 0.783 & 0.776 & 0.791 \\ 
  $\sigma_{1}$ & 0.021 & 0.805 & 0.894 & 0.955 & 0.027 & 0.805 & 0.894 & 0.955 & 0.835 & 0.000 & 0.000 & 0.000 \\ 
  $\sigma_{2}$ & 0.019 & 0.625 & 0.662 & 1.364 & 0.023 & 0.625 & 0.662 & 1.364 & 0.790 & 0.000 & 0.000 & 0.000 \\
\hline
\end{tabular}
\end{table}

Similar to scenario I, the estimation of the \gls{mfpc} basis mainly increases the variance of the estimates in the longitudinal submodel, with \gls{mse} values still lower than in the \textbf{JMbayes2} framework. The corresponding coverage drops to about $80\%$ for the longitudinal predictors and the variance predictors are overestimated with a coverage of $0$, both similar to the \textit{JMB} model. Figure \ref{fig:simII_modelfits} exemplifies that the longitudinal submodel fits are still sufficiently close to the true trajectories for the estimation of the association predictors $\eta_{\alpha_1}, \eta_{\alpha_2}$ not to suffer from the estimation of the \gls{mfpc} basis. Truncating the basis to $99\%$ of explained variance again entails a slight deterioration in the overall model fit as reported by the bias, \gls{mse}, and coverage. Note, however, that in $64\%$ of all simulation runs all six \glspl{mfpc} are included in the \textit{TRUNC} models. Further decreasing the truncation threshold to $95\%$ (results not shown) lowers the computational burden by reducing the number of \gls{mfpc} basis functions from six to on average four, thus dropping $2\times 300$ parameters. The resulting decline in estimation accuracy, however, is so considerable that the \gls{mse} values are then higher than for the oversmoothed \textit{JMB} models for both the longitudinal and the association predictors. We conclude that truncation needs to be carefully considered and that a large percentage of variation should be retained unless prohibited by computational considerations.

The time dependent evaluation in both simulation scenarios (results not shown) reveals that the estimation of the baseline hazard with P-splines in both frameworks leads to oversmoothing at the beginning of the time interval, where the baseline hazard is quickly changing (cf.\ Bender et al., 2018\cite{bender2018generalized}), explaining results in the first rows of Tables \ref{tab:simI_results} and \ref{tab:simII_results}. For the longitudinal submodel, the drop-out process implies less information at the end of the time interval, resulting in decreasing estimation accuracy. As the drop-out process also affects the estimation of the \gls{mfpc} basis functions and as this additional uncertainty is not reflected in the \textit{EST} and \textit{TRUNC} models, this also  results in decreasing coverage over time.  

The results from both simulation scenarios reinforce that the choice and for \glspl{mfpc} also estimation of the basis is an essential prerequisite for a satisfactory model fit. We see in scenario I that if the covariance structure is known to be parametric a priori, a parametric model will always perform better than our more flexible nonparametric specification, which by estimating the linear eigenfunctions introduces additional noise into the \glspl{mfpc}. On the other hand, parametric or less flexible spline specifications are not sufficient to capture  more complex trends and result in oversmoothing, such as in scenario II, where our more flexible approach has a clear advantage. We find that  estimated \glspl{mfpc} (see Appendix \ref{app:simulation}\ Figure \ref{APPfig:sim_scenII_mfpcs})
in most simulation runs recover the relevant characteristics of the eigenfunctions and thus time trends well, with the leading \glspl{mfpc} closer to the true eigenfunctions in simulation scenario II (cf.\ Table \ref{APPtab:mfpc_eval} in the Appendix). This illustrates again that the strength of our proposed \gls{mfpc} based approach lies in the estimation of complex covariance structures. For both simulation scenarios, we find that lowering the truncation threshold can considerably decrease the estimation accuracy of the longitudinal submodel and even negatively affect the survival submodel. We therefore advocate to use a high rate of explained variance, such as $99\%$, for determining the number of \glspl{mfpc} in a model.

\section{Application}
\label{sec:Application}

We demonstrate our parsimonious but flexible modeling approach by analyzing a \gls{pbc} study, where we want to model the association between different biochemical markers and the survival of the patients, and compare it to the more parameter intensive modeling approach of using spline based random effects to capture the (potentially) complex covariance structure in the data. The code to reproduce the analysis is provided in the accompanying \textsf{R} package \textbf{MJMbamlss}. The \gls{pbc} data provided in the \textsf{R} package \textbf{JMbayes2} include information on $312$ subjects, in particular on the follow-up time (composite event of liver transplantation or death), baseline covariates (sex, age, and treatment group), and several longitudinally measured biochemical outcomes which are suspected to be associated with disease progression. During the study period, visits were scheduled but missing observations and drop out lead to irregular measurements of the longitudinal outcomes with an average of $6.2$ visits per subject. At the end of the study period, $169$ subjects had experienced the event with median follow-up time of $4.5$ years (IQR: $2.2$, $6.5$) and $143$ subjects were event-free with median follow-up time of $8.7$ (IQR: $6.5$, $10.5$). For our analysis, we include the continuous longitudinal outcomes \textit{albumin} in g/dl, serum bilirubin (\textit{serBilir}) in mg/dl, serum cholesterol (\textit{serChol}) in mg/dl, and aspartate aminotransferase (\textit{SGOT}) in U/liter, all log-transformed to improve adherence to the normal assumption.\cite{pandolfi2023hidden} For a more direct comparison to the \textbf{JMbayes2} framework, eight subjects for whom no measurements of serum cholesterol are available are excluded from the analysis, giving a total of $304$ subjects. We use all available baseline information on the subjects such as \textit{age} at registration in years, \textit{sex} (male or female), and treatment group \textit{drug} (placebo or D-penicillamine) for modeling the hazard as well as the longitudinal outcomes. Note that we cap the follow-up times $t$ at the maximum observation time of longitudinal measurements in the data set ($14.1$ years, only affecting three event-free subjects) to avoid extrapolation of the longitudinal trajectories.

We apply the proposed \gls{mfpc} based approach (denoted as \textit{bamlss}) to the \gls{pbc} data and compare the results to a similarly specified model fit of the \textbf{JMbayes2} package (denoted as \textit{JMB}). These models follow the specifications \eqref{eq:hazard_one_observation} --  \eqref{eq:LongiPredictor} with 
\begin{align*}
    \eta_{\gamma i} &= \beta_{\gamma 0} + \beta_{\gamma 1}\cdot\text{sex}_i + \beta_{\gamma 2}\cdot\text{drug}_i + f_{\gamma}(\text{age}_i) \, , \\
    \eta_{\alpha_{k}i}(t) &= \beta_{\alpha_k} \, ,\\
    \eta_{\mu_{k}i}(t) &= \beta_{k0} + \beta_{k1}\cdot\text{sex}_{i} + \beta_{k2}\cdot\text{drug}_{i}+ f_{k1}(\text{age}_i) + f_{k2}(t) + b_i^{(k)}(t)\, , \, \text{and}\\
    \eta_{\sigma_k i}(t) &= \beta_{\sigma_k}
\end{align*}
for $k \in \{\textit{albumin}, \textit{serBilir}, \textit{serChol}, \textit{SGOT}\}$, and subject index $i$. To include a smooth time trend for the longitudinal outcomes and to allow for possibly nonlinear covariate effects of age, we use P-splines with ten basis functions before centering to estimate the additive functions $f_{\gamma}$, $f_{k1}$, and $f_{k2}$ with \textbf{bamlss}. For the \textbf{JMbayes2} model we try to fully exploit the framework's flexibility but for a natural cubic B-spline basis, we reach the limit with three basis functions, as more basis functions result in error messages. Similarly, we use the \gls{mfpc} based representation \eqref{eq:mfpc_represtation} to estimate the longitudinal random effects in our approach, while the \textbf{JMbayes2} model maximally allows random intercepts and a natural cubic splines representation of the follow-up time with two basis functions, giving three random effects per subject and longitudinal outcome. The baseline hazard $\eta_{\lambda i}(t)$ is modeled in both frameworks using cubic P-splines with second order difference penalty and 10 basis functions (before centering). 

For the estimation of the \gls{mfpc} basis, we follow the outline given in section \ref{subsec:EstimationMFPCA}. In order to obtain stable estimates of univariate scores, the \glspl{ufpca} are based on a subset of 145 subjects that have at least three longitudinal measurements and one measurement after $10\%$ of the follow-up time interval. We use seven marginal B-spline basis functions for the estimation of the univariate covariance functions and truncate the \gls{ufpca} at $99\%$ explained univariate variance, which results in three \glspl{ufpc} per longitudinal outcome. The resulting eigenvalues of the \glspl{ufpca} show a considerable imbalance in univariate variation in the data with marker \textit{serBilir} exhibiting an integrated univariate variance more than 70 times larger than that of \textit{albumin}. By using a scalar product \eqref{eq:scalarproduct} with weights inversely proportional to the integrated univariate variance for the \gls{mfpca}, we account for the different amounts of variation equivalent to standardizing the longitudinal data without actually transforming the outcome. The resulting \glspl{mfpc} are then more balanced across the longitudinal outcomes with the first two explaining $51.2\%$ and $22.8\%$ of the weighted variance, respectively, while using scalar product weights of $1$ would give a first \gls{mfpc} dominated by \textit{serBilir} accounting for $79.5\%$ of the (unweighted) multivariate variance. In a sensitivity analysis (see Appendix \ref{app:application}) we use the estimated \glspl{mfpc} of the unweighted \gls{mfpca} and find that the model fits better to the \textit{serBilir} data but worse to the other markers giving an overall worse model fit. Visual inspection of the estimated \glspl{mfpc} (see Appendix \ref{app:application} Figure \ref{APPfig:pbc_mpfcs}) shows that using seven B-spline basis functions in the univariate specifications yields a good compromise between overly smooth and too flexible (trailing) \glspl{mfpc}. To decrease the computational burden of the final model fit, we truncate the \gls{mfpc} basis at $99\%$ explained variance, which leads to nine included \glspl{mfpc} for the weighted \gls{mfpca} models, i.e.\ roughly a parsimonious two basis functions per marker.

\begin{figure}
    \centering
    \includegraphics[width=0.8\textwidth]{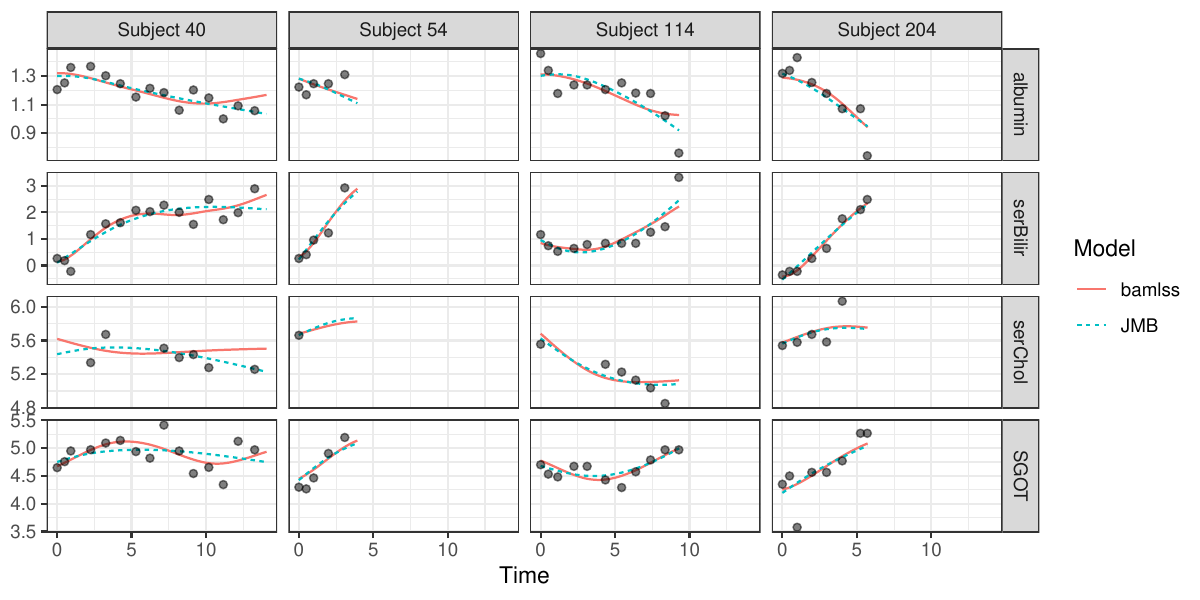}
    \caption{\gls{mfpc} based (\textit{bamlss}) and spline based (\textit{JMB}) model fits of the \gls{pbc} data for four subjects. The grey dots are the observed data points. Longitudinal outcomes are log-transformed.}
    \label{fig:pbc_modelfits}
\end{figure}

For each model, we run one \gls{mcmc} chain with 12,000 iterations, a burnin of 2,000 and a thinning of 5, yielding 2,000 samples. Overall, we find that the \gls{mfpc} based \textit{bamlss} model yields similar results to the spline based random effects approach of the \textit{JMB} model. In particular, Table \ref{tab:pbc_alpha} shows that we find significant association of the time-to-event process with the current value of each biochemical marker of our analysis, irrespective of the modeling approach used. This finding aligns with other works in the literature\cite{andrinopoulou2016bayesian,rustand2023fast,tu2023gaussian}, who might have come to slightly different conclusions due to different model specifications. Other estimated parameters such as the linear effects of \textit{sex} on the hazard or the longitudinal outcomes are also similar for the \textit{bamlss} and \textit{JMB} models (see Appendix \ref{app:application} Table \ref{APPtab:pbc_pars}). What is more, Figure \ref{fig:pbc_modelfits} illustrates that the longitudinal submodel fits of both models are generally very close. Given that the estimated \gls{mfpc} basis could provide much more flexibility than the spline approach with only two basis functions, this suggests that the PBC data might not require the added flexibility supplied by an \gls{mfpc} basis representation. Nevertheless, the \textit{bamlss} model uses much fewer parameters for the estimation of the random effects structure despite the larger flexibility. While the spline based approach estimates $3\times 4 \times 304$ random effects with an unstructured covariance matrix of $78$ covariance parameters, the \gls{mfpc} based approach only needs $9\times 304$ random effects with a diagonal covariance matrix of $9$ variance parameters, thus saving $3\times 304 + 69 = 981$ parameters total. We therefore show that using an \gls{mfpc} basis not only reduces the computational costs of fitting an \gls{mjm}, but also constitutes an alternative, highly flexible modelling approach to spline based random effects, especially when numerical problems restrict their usage for nonlinear covariance structures. Then the proposed approach also allows to check whether the assumptions made in a spline model with few basis functions dictated by computational possibilities are in fact warranted.

\begin{table}
\caption{Posterior mean and $95\%$ credible interval of the association predictors $\beta_{\alpha_k}$ for the two models also shown in Figure \ref{fig:pbc_modelfits}. Longitudinal outcomes are log-transformed.}
\label{tab:pbc_alpha}
\vspace{1em}
\footnotesize
\centering
\begin{tabular}{l|rr|rr}
\hline
& \multicolumn{2}{c|}{\textit{bamlss}} & \multicolumn{2}{c}{\textit{JMB}}\\
\hline
\textit{ albumin } &$ -5.82 $ & $[ -7.84 $; $ -3.89 ]$ &  $ -6.03 $ & $[ -7.87 $;  $ -4.24 ]$ \\  \textit{ serBilir } &$ 1.48 $ & $[ 1.19 $; $ 1.77 ]$ &  $ 1.39 $ & $[ 1.11 $;  $ 1.68 ]$ \\  \textit{ serChol } &$ -0.76 $ & $[ -1.36 $; $ -0.16 ]$ &  $ -0.68 $ & $[ -1.26 $;  $ -0.06 ]$ \\  \textit{ SGOT } &$ -0.74 $ & $[ -1.41 $; $ -0.13 ]$ &  $ -0.76 $ & $[ -1.4 $;  $ -0.11 ]$ \\ 
\hline
\end{tabular}
\end{table}

\section{Discussion and Outlook}
\label{sec:Discussion}

We have shown that shared parameter \glspl{mjm} can greatly benefit from using an \gls{mfpc} based representation of the random effects structure. In particular, the flexibility when modeling highly nonlinear subject specific trends in the multivariate longitudinal outcomes is often restricted by the computational complexity of the number of parameters to estimate, especially when a spline approach is used. Using \glspl{mfpc} as a basis for random effects allows a parsimonious representation of increased flexibility, reducing the number of parameters to estimate by decoupling it from the number of longitudinal outcomes and simplifying the covariance structure of the random effects. Our simulation study showcases that the \gls{mfpc} based approach can achieve a better model fit for nonlinear longitudinal outcome trajectories resulting also in a more accurate estimation of the association structure with the event hazard. Furthermore, the explicit ordering of the \glspl{mfpc} in terms of explained variance makes the trade-off between computational burden and goodness of fit explicit, which allows its steering by the selection of the truncation level for the number of \gls{mfpc} based random effects.
Note that while not focused on here, dynamic prediction for subjects with incoming new longitudinal measurements is directly available from our approach since the estimated \glspl{mfpc} are used as empirical basis functions also for all (new) subjects in the proposed \gls{mjm}, and the standard algorithms for dynamic prediction of the random scores can thus be used.\cite{rizopoulos2016r}

We also provide a working implementation for the framework of flexible \glspl{mjm} in the \textbf{bamlss} package, which supports a vast range of possible specifications for covariate effects in a user friendly way. Note, however, that the idea of an \gls{mfpc} basis is more universal and can be combined with different approaches to shared parameter \glspl{mjm} such as approximations using adapted two-stage models\cite{mauff2020joint} or \gls{inl}\cite{rustand2023fast}. For \textbf{JMbayes2}, both the simulation study and our application have shown that the model flexibility is restricted by the number of spline basis functions which can be included in the longitudinal submodel. This could be mitigated by implementing our proposed \gls{mfpc} approach also in \textbf{JMbayes2}. Additionally, Rustand et al. (2020)\cite{rustand2023fast} remark that simplifying the random effects covariance matrix to be block-diagonal considerably reduces computation times. While this corresponds to the strong assumption of independent random effects across markers, an \gls{mfpc} basis does in fact provide the further simplification of a diagonal covariance matrix without the necessity of such an independence assumption. On the other hand, both  the adapted two-stage and the INL approximations to the full \gls{mjm} present computational advantages to the derivative based \gls{mcmc} sampling implemented in \textbf{bamlss}. In simulation scenario I, for example, the fully Bayesian approach in \textbf{bamlss} estimates the \gls{mjm} in around $30$ minutes on a Linux system with $12$ cores at $3.5$ GHz and $32$ GB memory while the highly optimized approach in \textbf{JMbayes2} returns results after roughly one minute with the same setup. This difference might in future be diminished by increasing computational efficiency in the implemented code via transfering more computations to other languages such as \textsf{C} or exploiting sparse structures in the design matrices. A combination of an \gls{mfpc} approach and approximate approaches also holds promise.

It is important to note, however, that estimating the \gls{mfpc} basis is an integral part of our proposed model estimation approach. Besides model selection criteria such as the Bayesian information criterion or the deviance information criterion, we find that examination of the amount of variance explained and visual inspection of the estimated \glspl{ufpc} and \glspl{mfpc} is helpful to select the number of \glspl{mfpc} and assure that the fitted longitudinal trajectories in the \gls{mjm} exhibit a reasonable amount of flexibility. Our current approach of estimating the \gls{mfpc} basis ignores the informative missingness of the longitudinal process by treating the longitudinal outcomes as sparsely observed functional data. While we are able to show that the resulting estimated \glspl{mfpc} serve as a sufficiently good empirical basis in the \gls{mjm}, we encourage further research on estimating \glspl{ufpc} and \glspl{mfpc} under informative censoring. Several ad-hoc adjustments might improve the efficient use of all available longitudinal information such as weighting univariate outcomes depending on their length of follow-up in the estimation of the univariate covariance structure instead of excluding subjects with short follow-up time from the \gls{mfpca} entirely. Dong et al. (2023)\cite{dong2023jointly} suggest that the \gls{ufpc} estimation under informative censoring might be improved by iteratively estimating \glspl{ufpc} and predicting the full longitudinal trajectories, as this might stabilize estimation. Alternatively, separate univariate joint models based on reduced rank models, which simultaneously estimate the time-to-event process and the \glspl{ufpc} as in Yao (2007)\cite{yao2007functional}, could be employed, but would considerably increase computational burden. Further research is needed to evaluate and compare different approaches to estimate \glspl{ufpc} under informative missingness.

As a consequence of the proposed two step approach, all model based inference is conditional on the estimated \glspl{mfpc}. Since Happ and Greven \cite{happ2018multivariate} show that the \gls{mfpc} estimates asymptotically converge to the true eigenfunctions, the inference can be expected to be asymptotically correct. Future work is needed to account for the uncertainty in the estimation of the \gls{mfpc} basis in finite samples, for example using corrections along the lines of Goldsmith et al. (2013)\cite{goldsmith2013corrected}, which would improve the frequentist coverage properties of the longitudinal submodel fits. Developing an estimation approach which jointly estimates the scores and \glspl{mfpc} within an \gls{mjm} might further enhance modeling and understanding of the longitudinal processes, also allowing for inference regarding the estimated \glspl{mfpc}.

Given that an \gls{mfpc} based representation of random effects is a universal and effective approach for \glspl{mjm}, future work can follow several more research avenues. From extending the presented joint model to incorporate functional covariates, functional outcomes, or different parameterizations of the association structure between the longitudinal and time-to-event submodels, to developing efficient approximations to the full derivative based \gls{mcmc} algorithm in our \textbf{bamlss} implementation or advancing the \gls{mfpc} basis approach to non-normal longitudinal outcomes, \glspl{mjm} remain an interesting and challenging field of study.

\section{Acknowledgements} 
We gratefully acknowledge funding by grant GR 3793/3-1 `Flexible regression methods for curve and shape data' from the German Research Foundation (DFG).

\bibliographystyle{abbrvnat}
\bibliography{lit}

\appendix

\section{Derivatives}
\label{app:derivatives}

In the following, the blockwise score vectors $\bm{s}(\bm{\beta}_{lh})$ and Hessian $\bm{\mathcal{H}}(\bm{\beta}_{lh})$ used in the posterior mode and mean estimation are presented. Note that we show only the derivatives based on the log-likelihood function as adding the derivatives for the respective log-priors is straightforward.\cite{kohler2017flexible} To highlight the differences between the marker-specific fixed and principal component based random effects, we denote design matrices containing evaluations of the fixed effects as $\bm{X}_{\mu}$ and those containing evaluations of the \gls{mfpc} basis as $\bm{\Psi}$. The $u$-th row of a design matrix is given as $\bm{x}_{lu}^{\top}$ and $\bm{\psi}_u^{\top}$, respectively. The additional index $(m)$ for a \gls{mfpc} basis matrix $\bm{\Psi}_{(m)}$ or a vector $\bm{\psi}_{(m)}$ denotes the submatrix of $\bm{\Psi}$ or the subvector of $\bm{\psi}$ containing only columns or elements corresponding to the $m$-th \gls{mfpc}. 

For the survival submodel, $\bm{X}_{\gamma}$ is the  $n\times d_\gamma$ design matrix for the baseline covariates. The $n\times d_l$ design matrix $\bm{X}_l(\bm{T}), l \in \{\lambda, \alpha_1,...,\alpha_K, \mu_1, ...,\mu_K\}$ contains evaluations of the corresponding $\tilde{H}_l$ model terms at the event times $\bm{T}$, where here and in the following bold arguments of matrices stand for row-wise argument evaluation. The principal component based random effect design matrix $\bm{\Psi}^{(k)}(\bm{T}) = blockdiag(\bm{\Psi}_1^{(k)}(T_1),..., \bm{\Psi}_n^{(k)}(T_n))$ is blockdiagonal with blocks of size $1\times M$. 

For the longitudinal submodel, the $N_k\times d_l$ design matrix $\bm{X}_l(\bm{t}_k), l \in \{\mu_k, \sigma_k| k = 1,...,K\}$ and the marker-specific $N_k\times Mn$ blockdiagonal matrix $\bm{\Psi}^{(k)}(\bm{t}_k) = blockdiag(\bm{\Psi}_1^{(k)}(\bm{t}_{1k}),..., \bm{\Psi}_n^{(k)}(\bm{t}_{nk}))$ contain evaluations of the corresponding $\tilde{H}_l$ model terms and the \gls{mfpc} based functional random effects at the $N_k = \sum_i n_{ik}$ repeated measurement times $\bm{t}_k = (\bm{t}_{1k}^{\top},..., \bm{t}_{nk}^{\top})^{\top}$ of individual and outcome specific observation times $\bm{t}_{ik} = (t_{ik1}, ..., t_{ikn_{ik}})^{\top}$, respectively. We also define the fixed effects design matrix of the full multivariate longitudinal submodel $\bm{X}_{\mu}(\bm{t}) = blockdiag(\bm{X}_{\mu_1}(\bm{t}_1),..., \bm{X}_{\mu_K}(\bm{t}_K))$ comprising all longitudinal observations at respective time-points $\bm{t} = (\bm{t}^{\top}_1,...,\bm{t}_K^{\top})^{\top}$. The corresponding $N\times Mn$ random effect matrix $\bm{\Psi}(\bm{t})$ stacks the marker-specific $N_k\times Mn$  matrices. Analogously to $\bm{X}_\mu(\bm{t})$, the design matrix for the variance of the Gaussian noise $\bm{X}_{\sigma}(\bm{t}) = blockdiag(\bm{X}_{\sigma_1}(\bm{t}_1),..., \bm{X}_{\sigma_K}(\bm{t}_K))$ is blockdiagonal with $N_k\times d_{\sigma_k}$ matrix $\bm{X}_{\sigma_k}(\bm{t}_k)$ and we define $\bm{R} = blockdiag(\bm{R}_{1},..., \bm{R}_K)$ with $\bm{R}_k = diag(\exp(\bm{X}_{\sigma_k}(\bm{t}_k)\bm{\beta}_{\sigma_k})^2)$. The corresponding parameter vectors $\bm{\beta}_{\mu} = (\bm{\beta}_{\mu_1}^{\top}, ..., \bm{\beta}_{\mu_K}^{\top})^{\top}$ and $\bm{\beta}_{\sigma} = (\bm{\beta}_{\sigma_1}^{\top}, ..., \bm{\beta}_{\sigma_K}^{\top})^{\top}$ are stacked over longitudinal outcomes, while the random score vector $\bm{\rho} = (\bm{\rho}_1^{\top},..., \bm{\rho}_n^{\top})^{\top}$ is stacked over subjects.

The log-likelihood of the \gls{mjm} can be written as
\begin{align*}
\ell(\bm{\theta} \mid  \bm{T}, \bm{ \delta}, \bm{y}) = &-\frac{N}{2}\log(2\pi) - \bm{1}_{N}^{\top}\bm{X}_{\sigma}(\bm{t})\bm{\beta}_{\sigma} - \frac{1}{2}\left(\bm{y}- \bm{X}_{\mu}(\bm{t})\bm{\beta}_{\mu} - \bm{\Psi}\left(\bm{t})\bm{\rho}\right)^{\top}\bm{R}^{-1}(\bm{y}- \bm{X}_{\mu}(\bm{t})\bm{\beta}_{\mu} - \bm{\Psi}(\bm{t})\bm{\rho}\right) \\
&+ \bm{\delta}^{\top}\bigg(\bm{X}_{\lambda}(\bm{T}) \bm{\beta}_\lambda + \bm{X}_{\gamma}\bm{\beta}_\gamma + \sum_{k= 1}^{K}\big(\bm{X}_{\alpha_k}(\bm{T})\bm{\beta}_{\alpha_k}\big)\cdot \big(\bm{X}_{\mu_k}(\bm{T})\bm{\beta}_{\mu_k} + \bm{\Psi}^{(k)}(\bm{T})\bm{\rho}\big)\bigg)\\
&- \sum_{i=1}^{n}\exp(\bm{x}_{\gamma i}^{\top}\bm{\beta}_{\gamma})\int_{s=0}^{T_i}\exp\bigg(\bm{x}_{\lambda i}(s)^{\top}\bm{\beta}_{\lambda} + \sum_{k = 1}^{K}\big(\bm{x}_{\alpha_k i}(s)^{\top}\bm{\beta}_{\alpha_k}\big)\big(\bm{x}_{\mu_k i}(s)^{\top}\bm{\beta}_{\mu_k} + \bm{\psi}_{i}^{(k)}(s)^{\top}\bm{\rho}_i\big)\bigg)ds
\end{align*}
with $\cdot$ denoting elementwise multiplication of vectors. For notational ease, write $w_i(s) = \exp\big(\bm{x}_{\lambda i}(s)^{\top}\bm{\beta}_{\lambda} + \sum_{k = 1}^{K}\big(\bm{x}_{\alpha_k i}(s)^{\top}\bm{\beta}_{\alpha_k}\big)\cdot \big(\bm{x}_{\mu_k i}(s)^{\top}\bm{\beta}_{\mu_k} + \bm{\psi}_{i}^{(k)}(s)^{\top}\bm{\rho}_i\big)\big)$.

Then the score vectors of the log-likelihood function are given as
\begin{align*}
\bm{s}(\bm{\beta}_{\mu_k}) = \frac{\partial \ell}{\partial\bm{\beta}_{\mu_k}} & =  \bm{X}_{\mu_k}(\bm{t}_k)^{\top}\bm{R}_{k}^{-1}\left(\bm{y}_k - \bm{X}_{\mu_k}(\bm{t}_k)\bm{\beta}_{\mu_k} - \bm{\Psi}^{(k)}(\bm{t}_k)\bm{\rho}\right) + \bm{X}_{\mu_k}(\bm{T})^{\top}diag(\bm{\delta})\bm{X}_{\alpha_k}(\bm{T})\bm{\beta}_{\alpha_k}\\
& \quad -\sum_{i = 1}^{n} \exp(\bm{x}_{\gamma i}^{\top}\bm{\beta}_{\gamma})\int_{0}^{T_i}  w_i(s) \bm{x}_{\alpha_k i}(s)^{\top} \bm{\beta}_{\alpha_k} \cdot \bm{x}_{\mu_k i}(s) ds\\
\bm{s}(\bm{\rho}_{(m)}) = \frac{\partial \ell}{\partial\bm{\rho}_{(m)}} &=\bm{\Psi}_{(m)}(\bm{t})^{\top}\bm{R}^{-1}\left(\bm{y} - \bm{X}_{\mu}(\bm{t})\bm{\beta}_{\mu} - \bm{\Psi}(\bm{t})\bm{\rho}\right) + \sum_{k=1}^{K}\bm{\Psi}_{(m)}^{(k)}(\bm{T})diag(\bm{\delta})\bm{X}_{\alpha_k}(\bm{T})\bm{\beta}_{\alpha_k}\\
& \quad - \sum_{i = 1}^{n} \exp(\bm{x}_{\gamma i}^{\top}\bm{\beta}_{\gamma i})\int_{0}^{T_i}  w_i(s)\bigg(\sum_{k=1}^{K}\big(\bm{x}_{\alpha_k i}(s)^{\top} \bm{\beta}_{\alpha_k}\big)\cdot \bm{\psi}_{(m)i}^{(k)}(s)\bigg) ds\\
\bm{s}(\bm{\beta}_{\alpha_k})  = \frac{\partial \ell}{\partial\bm{\beta}_{\alpha_k}}  &= \bm{X}_{\alpha_k}(\bm{T})^{\top}diag(\bm{\delta})\left(\bm{X}_{\mu_k}(\bm{T})\bm{\beta}_{\mu_k} + \bm{\Psi}^{(k)}(\bm{T})\bm{\rho}\right) \\
& \quad - \sum_{i = 1}^{n}\exp(\bm{x}_{\gamma i}^{\top}\bm{\beta}_{\gamma i})\int_{0}^{T_i}  w_i(s) \big( \bm{x}_{\mu_k i}(s)^{\top}\bm{\beta}_{\mu_k} + \bm{\psi}_i^{(k)}(s)^{\top}\bm{\rho}_i\big) \cdot \bm{x}_{\alpha_k i}(s)ds\\
\bm{s}(\bm{\beta}_{\lambda}) = \frac{\partial \ell}{\partial\bm{\beta}_{\lambda}} &= \bm{X}_{\lambda}(\bm{T})^{\top}\bm{\delta} - \sum_{i = 1}^{n} \exp(\bm{x}_{\gamma_i}^{\top}\bm{\beta}_{\gamma})\int_{0}^{T_i}w_i(s) \bm{x}_{\lambda i}(s)ds\\
\bm{s}(\bm{\beta}_{\gamma}) = \frac{\partial \ell}{\partial\bm{\beta}_{\gamma}}&= \bm{X}_{\gamma}^{\top}\bm{\delta} - \sum_{i = 1}^{n} \exp(\bm{x}_{\gamma_i}^{\top}\bm{\beta}_{\gamma})\bm{x}_{\gamma_i}\int_{0}^{T_i}w_i(s)ds\\
\bm{s}(\bm{\beta}_{\sigma})  = \frac{\partial \ell}{\partial\bm{\beta}_{\sigma}}& = - \bm{X}_{\sigma}(\bm{t})^{\top}\bm{1}_{N} + \bigg(\bm{X}_{\sigma}(\bm{t}) \odot \big(\bm{y} - \bm{X}_{\mu}(\bm{t})\bm{\beta}_{\mu} - \bm{\Psi}(\bm{t})\bm{\rho}\big)\bigg)^{\top}\bm{R}^{-1}\big(\bm{y} - \bm{X}_{\mu}(\bm{t})\bm{\beta}_{\mu} - \bm{\Psi}(\bm{t})\bm{\rho}\big)
\end{align*}
with row tensor product $\odot$ of a $p \times a$ matrix $A$ and a $p \times b$ matrix $B$ defined as the $p\times ab$ matrix $A\odot B = (A\otimes \bm{1}_b^{\top}) \cdot (\bm{1}_a^{\top} \otimes B)$ and $\otimes$ the Kronecker product.

The corresponding Hessians are 
\begin{align*}
\bm{\mathcal{H}}(\bm{\beta}_{\mu_k}) &= \frac{\partial \ell}{\partial\bm{\beta}_{\mu_k}\partial\bm{\beta}_{\mu_k}^{\top}} = -\bm{X}_{\mu_k}(\bm{t}_k)^{\top}\bm{R}_{k}^{-1}\bm{X}_{\mu_k}(\bm{t}_k) - \sum_{i = 1}^{n} \exp(\bm{x}_{\gamma i}^{\top}\bm{\beta}_{\gamma i})\int_{0}^{T_i}  w_i(s) \big(\bm{x}_{\alpha_k i}(s)^{\top} \bm{\beta}_{\alpha_k}\big)^2 \cdot \bm{x}_{\mu_k i}(s)\bm{x}_{\mu_k i}(s)^{\top} ds\\
\bm{\mathcal{H}}(\bm{\rho}_{(m)}) &= \frac{\partial \ell}{\partial\bm{\rho}_{(m)}\partial\bm{\rho}_{(m)}^{\top}} = -\bm{\Psi}_{(m)}(\bm{t})^{\top}\bm{R}^{-1}\bm{\Psi}_{(m)}(\bm{t}) \\& \qquad\qquad\qquad\quad - \sum_{i = 1}^{n} \exp(\bm{x}_{\gamma i}^{\top}\bm{\beta}_{\gamma i})\int_{0}^{T_i}  w_i(s) \bigg(\sum_{k=1}^{K}\big(\bm{x}_{\alpha_k i}(s)^{\top} \bm{\beta}_{\alpha_k}\big)\cdot \bm{\psi}_{(m)i}^{(k)}(s)\bigg)\bigg(\sum_{k=1}^{K}\big(\bm{x}_{\alpha_k i}(s)^{\top} \bm{\beta}_{\alpha_k}\big)\cdot \bm{\psi}_{(m)i}^{(k)}(s)\bigg)^{\top}ds\\
\bm{\mathcal{H}}(\bm{\beta}_{\alpha}) &= \frac{\partial \ell}{\partial\bm{\beta}_{\alpha_k}\partial\bm{\beta}_{\alpha_k}^{\top}} = - \sum_{i = 1}^{n}\exp(\bm{x}_{\gamma i}^{\top}\bm{\beta}_{\gamma i})\int_{0}^{T_i}  w_i(s) \big(\bm{x}_{\mu_k i}(s)^{\top}\bm{\beta}_{\mu_k} + \bm{\psi}_i^{(k)}(s)^{\top}\bm{\rho}_i\big)^2 \cdot \bm{x}_{\alpha_k i}(s)\bm{x}_{\alpha_k i}(s)^{\top} ds\\
\bm{\mathcal{H}}(\bm{\beta}_{\lambda}) &= \frac{\partial \ell}{\partial\bm{\beta}_{\lambda}\partial\bm{\beta}_{\lambda}^{\top}} = - \sum_{i = 1}^{n} \exp(\bm{x}_{\gamma_i}^{\top}\bm{\beta}_{\gamma})\int_{0}^{T_i}w_i(s) \bm{x}_{\lambda i}(s)  \bm{x}_{\lambda i}(s)^{\top}ds\\
\bm{\mathcal{H}}(\bm{\beta}_{\gamma}) &= \frac{\partial \ell}{\partial\bm{\beta}_{\gamma}\partial\bm{\beta}_{\gamma}^{\top}} = - \sum_{i = 1}^{n} \exp(\bm{x}_{\gamma_i}^{\top}\bm{\beta}_{\gamma})\bm{x}_{\gamma_i}\bm{x}_{\gamma_i}^{\top}\int_{0}^{T_i}w_i(s)ds\\
\bm{\mathcal{H}}(\bm{\beta}_{\sigma}) &= \frac{\partial \ell}{\partial\bm{\beta}_{\sigma}\partial\bm{\beta}_{\sigma}^{\top}}  =- 2\bigg(\bm{X}_{\sigma}(\bm{t}) \odot \big(\bm{y} - \bm{X}_{\mu}(\bm{t})\bm{\beta}_{\mu} - \bm{\Psi}(\bm{t})\bm{\rho}\big)\bigg)^{\top}\bm{R}^{-1}\bigg(\bm{X}_{\sigma}(\bm{t}) \odot \big(\bm{y} - \bm{X}_{\mu}(\bm{t})\bm{\beta}_{\mu} - \bm{\Psi}(\bm{t})\bm{\rho}\big)\bigg).
\end{align*}

\section{Simulation}
\label{app:simulation}

We present details for the two simulation scenarios such as the used parameter specifications and computational information. The code to reproduce the simulation is provided in the accompanying \textsf{R} package \textbf{MJMbamlss}.

\subsection{Description Scenario I}

The data generating process of Scenario I is based on the following additive predictors:
\begin{align*}
\eta_{\lambda i}(t) &=  1.37 \cdot t^{0.37}\\
\eta_{\gamma i} &= -1.5 + 0.48\cdot x_i\\
\eta_{\mu_k i}(t) &= 0.2\cdot t - 0.25 \cdot x_i - 0.05 \cdot t \cdot x_i +
                           b_{k1i} + b_{k2i} \cdot t ,
\\
\eta_{\alpha_1 i}(t) &= 
1.5, \, \eta_{\alpha_2 i}(t) = 0.6, \, \eta_{\alpha_3 i}(t) = 0.3 , \, \eta_{\alpha_4 i}(t) = -0.3, \, \eta_{\alpha_5 i}(t) = -0.6, \, \eta_{\alpha_6 i}(t) = -1.5\\
\eta_{\sigma_k i}(t) &= \log(0.06),\quad k = 1,..., 6
\end{align*}
where the random effects $\bm{b}_i = (b_{11i}, b_{12i}, b_{21i},..., b_{61i}, b_{62i})^{\top}$ are independent draws from a 12-dimensional multivariate normal $\mathcal{N}_{12}(\bm{0}, \bm{\Sigma})$ with
\begin{align*}
\bm{\Sigma} &= \begin{pmatrix} \bm{\Sigma}_1 & \bm{\Sigma}_2^\top \\
\bm{\Sigma}_2 & \bm{\Sigma}_3
\end{pmatrix}
\end{align*}
and
\begin{gather*}
\bm{\Sigma}_1 = \begingroup \setlength\arraycolsep{2pt} \begin{pmatrix}
   0.080 & -0.070 & 0.030 & 0.030 & 0.022 & 0.022 \\
    -0.070 & 0.900 & 0.030 & 0.030 & 0.022 & 0.022 \\
    0.030 & 0.030 & 0.096 & -0.084 & 0.030 & 0.030 \\
    0.030 & 0.030 & -0.084 & 1.080 & 0.030 & 0.030 \\
    0.022 & 0.022 & 0.030 & 0.030 & 0.112 & -0.098 \\
    0.022 & 0.022 & 0.030 & 0.030 & -0.098 & 1.260 
\end{pmatrix}\endgroup, \quad
\bm{\Sigma}_2 = 
\begingroup \setlength\arraycolsep{2pt} \begin{pmatrix}
    0.015 & 0.015 & 0.022 & 0.022 & 0.030 & 0.030 \\
    0.015 & 0.015 & 0.022 & 0.022 & 0.030 & 0.030 \\
    0.000 & 0.000 & 0.015 & 0.015 & 0.022 & 0.022 \\
    0.000 & 0.000 & 0.015 & 0.015 & 0.022 & 0.022 \\
    0.000 & 0.000 & 0.000 & 0.000 & 0.015 & 0.015 \\
    0.000 & 0.000 & 0.000 & 0.000 & 0.015 & 0.015 \\
\end{pmatrix} \endgroup, \\
 \bm{\Sigma}_3 =  \begingroup\setlength\arraycolsep{2pt}\begin{pmatrix}
     0.128 & -0.112 & 0.030 & 0.030 & 0.022 & 0.022 \\ 
   -0.112 & 1.440 & 0.030 & 0.030 & 0.022 & 0.022 \\ 
   0.030 & 0.030 & 0.144 & -0.126 & 0.030 & 0.030 \\ 
  0.030 & 0.030 & -0.126 & 1.620 & 0.030 & 0.030 \\
  0.022 & 0.022 & 0.030 & 0.030 & 0.160 & -0.140 \\ 
  0.022 & 0.022 & 0.030 & 0.030 & -0.140 & 1.800 \\ 
 \end{pmatrix}\endgroup.
\end{gather*}

\subsection{Description Scenario II}

\begin{figure}
    \centering
    \includegraphics[width=0.8\textwidth]{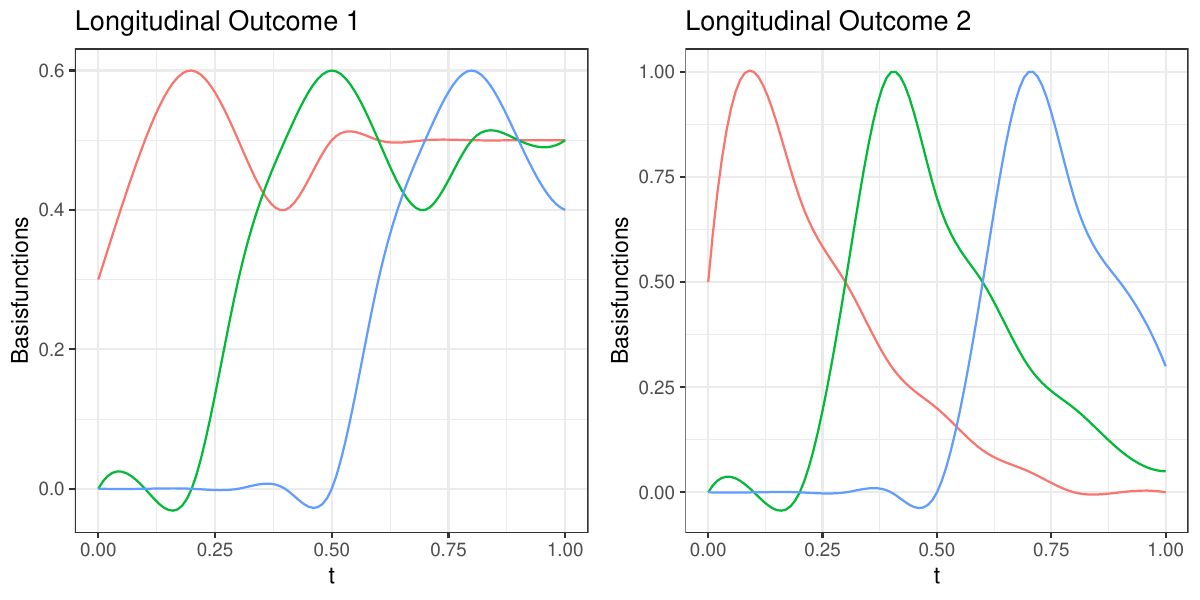}
    \caption{Non-orthogonal univariate basis functions used in simulation scenario II to generate a multivariate covariance kernel based on a finite \gls{kl} decomposition.}
    \label{APPfig:sim_basis_fcts}
\end{figure}

\begin{figure}
    \centering
    \includegraphics[width=0.8\textwidth]{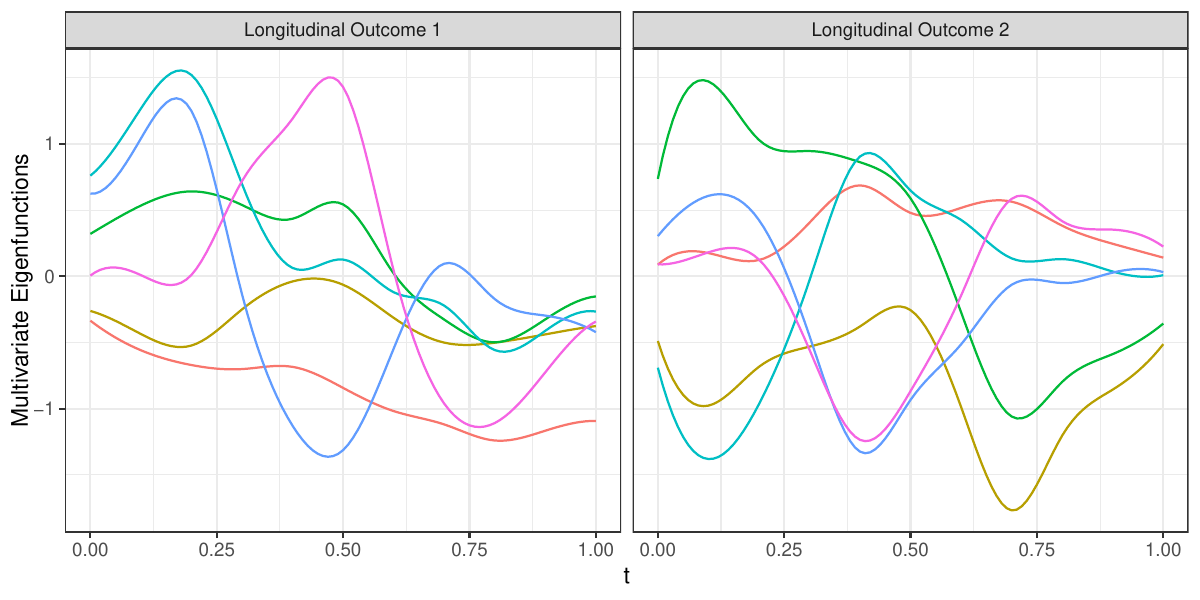}
    \caption{Multivariate eigenfunctions of the covariance kernel used in simulation scenario II. Colors correspond to the same multivariate functional principal components, $\psi_m^{(1)}$ and $\psi_m^{(2)}$.}
    \label{APPfig:sim_eigenfcts}
\end{figure}

The data generating process of Scenario II is based on the following additive predictors:
\begin{align*}
    \eta_{\lambda i}(t) &=  1.65 \cdot t^{0.65}\\
\eta_{\gamma i} &= -3 + 0.3 \cdot x_i\\
\eta_{\mu_k i}(t) &= 1\cdot t + 0.3 \cdot x_i + 0.3 \cdot t \cdot x_i + \sum_{m = 1}^6\rho_{im}\psi_{m}^{(k)}(t)
\\
\eta_{\alpha_k i}(t) &= 
1.1\\
\eta_{\sigma_k i}(t) &= \log(0.06), \quad k = 1, 2.
\end{align*}
The covariance structure induced by $\sum_{m = 1}^6\rho_{im}\psi_{m}^{(k)}(t)$ is based on a Gaussian process with covariance kernel $\mathcal{K}$ whose eigenfunctions are constructed solving the eigenanalysis problem 
\begin{align*}
    \bm{B}\bm{Q}\bm{c} = \nu \bm{c}
\end{align*}
with eigenvector $\bm{c}$, eigenvalue $\nu$, block diagonal matrix $\bm{B}$ of scalar products of the univariate basis functions given in Figure \ref{APPfig:sim_basis_fcts}, and the covariance matrix $\bm{Q}$ of the weights of the univariate basis expansion. The basis functions in Figure \ref{APPfig:sim_basis_fcts} reflect changes in the trajectories at the beginning, middle, or end of the time interval. $\bm{Q}$ is generated as a random covariance matrix with (rounded to three digits)
\begin{align*}
    \bm{Q} = \begin{pmatrix}
        3.124 & -0.396 & 0.892 & 0.119 & -0.668 & 0.005 \\ 
  -0.396 & 1.657 & 0.162 & -0.265 & -0.495 & -0.778 \\ 
  0.892 & 0.162 & 1.980 & -0.015 & -0.906 & 0.491 \\ 
  0.119 & -0.265 & -0.015 & 1.081 & 0.063 & 0.728 \\ 
  -0.668 & -0.495 & -0.906 & 0.063 & 0.890 & 0.243 \\ 
  0.005 & -0.778 & 0.491 & 0.728 & 0.243 & 1.969 
    \end{pmatrix}.
\end{align*} 
The resulting eigenvectors $\hat{\bm{c}}$ and eigenvalues given as (rounded to three digits) $\nu_1 = 1.376$, $\nu_2 = 0.531$, $\nu_3 = 0.149$, $\nu_4 = 0.101$, $\nu_5 = 0.044$, and $\nu_6 = 0.019$ are then used to calculate the corresponding multivariate eigenfunctions $\bm{\psi}_m, m = 1,..., 6$ (cf.\ Happ and Greven, 2018, Section 3.2\cite{happ2018multivariate}). The resulting multivariate eigenfunctions are given in Figure \ref{APPfig:sim_eigenfcts} and the scores $\rho_{im}$ are independent draws from $N(0, \nu_m)$.

\subsection{Additional Results}

\begin{figure}
    \centering \includegraphics[width=\textwidth]{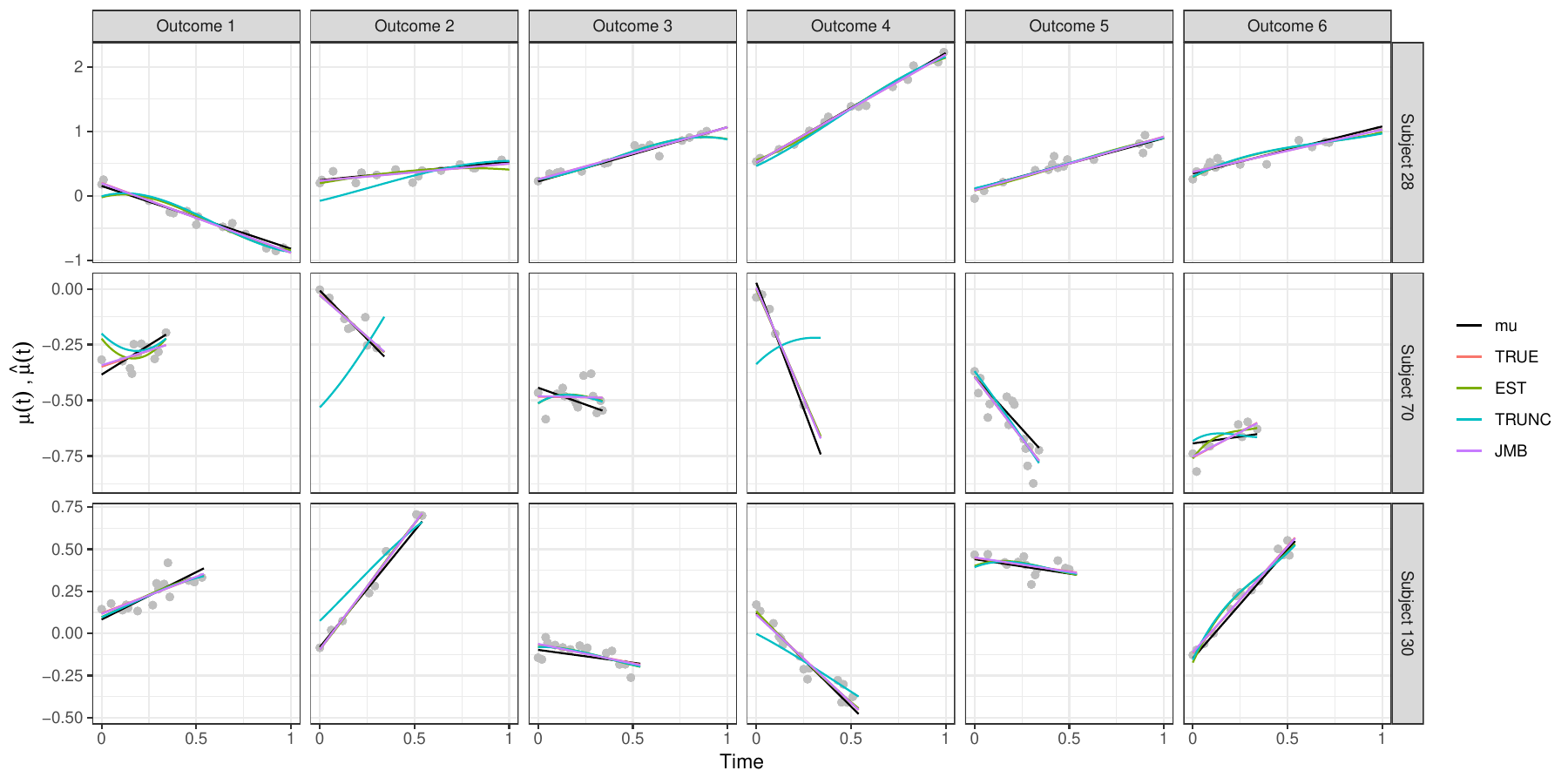}
    \caption{Exemplary model fits to the observed longitudinal data (grey dots) of three random subjects in one simulation iteration for simulation scenario I. The black lines (mu) correspond to the true trajectories.}
    \label{APPfig:sim_scenI_fits}
\end{figure}

\begin{figure}
    \centering \includegraphics[width=\textwidth]{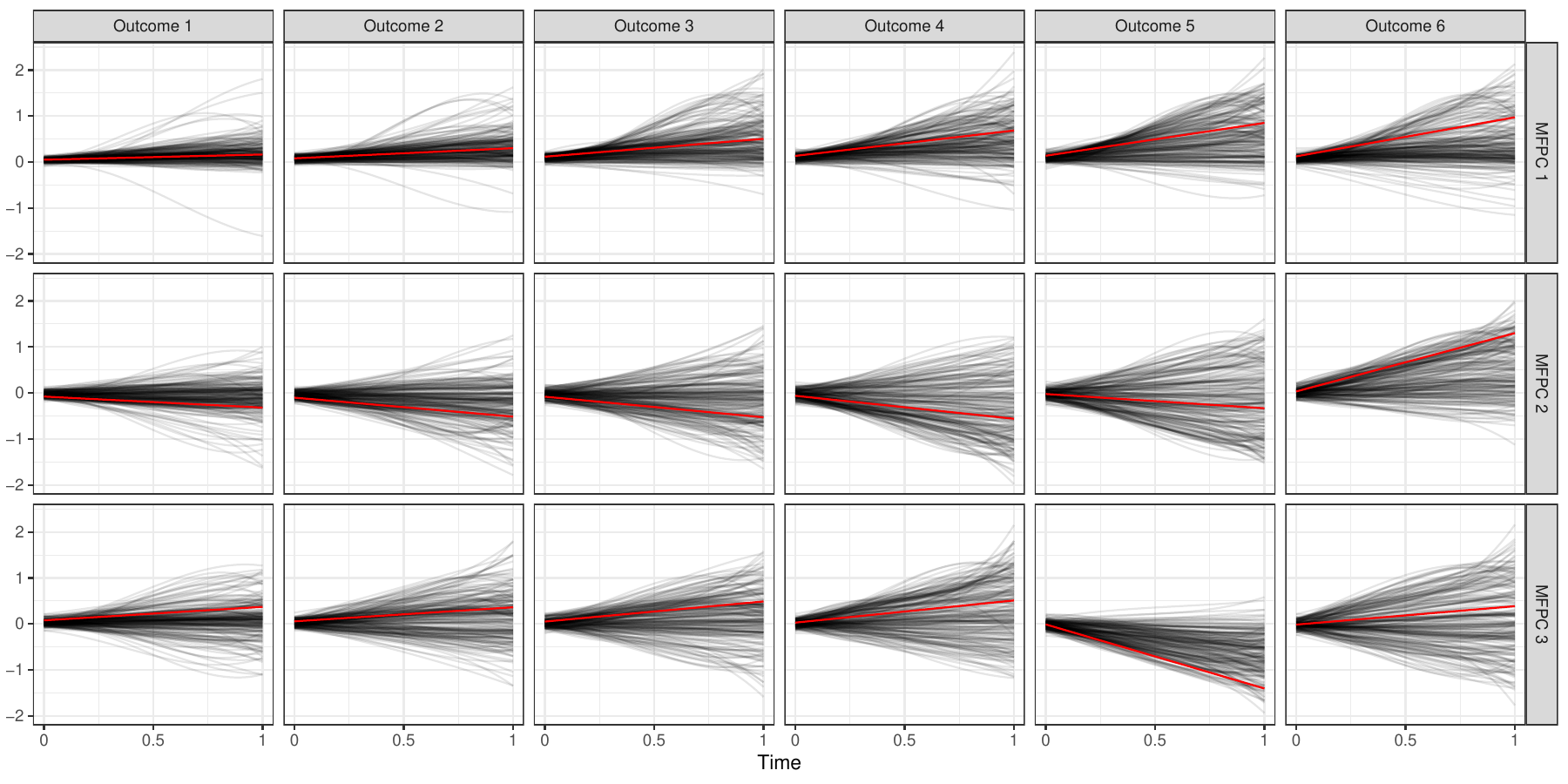}
    \caption{Estimated \gls{mfpc} basis functions $\hat{\bm{\psi}}_m(t)$ for the first three eigenfunctions (red) in simulation scenario I. Estimates are flipped if a reflected version $-\hat{\bm{\psi}}_m(t)$ is closer to the true $\bm{\psi}_m(t)$.}
    \label{APPfig:sim_scenI_mfpcs}
\end{figure}

\begin{figure}
    \centering \includegraphics[width=\textwidth]{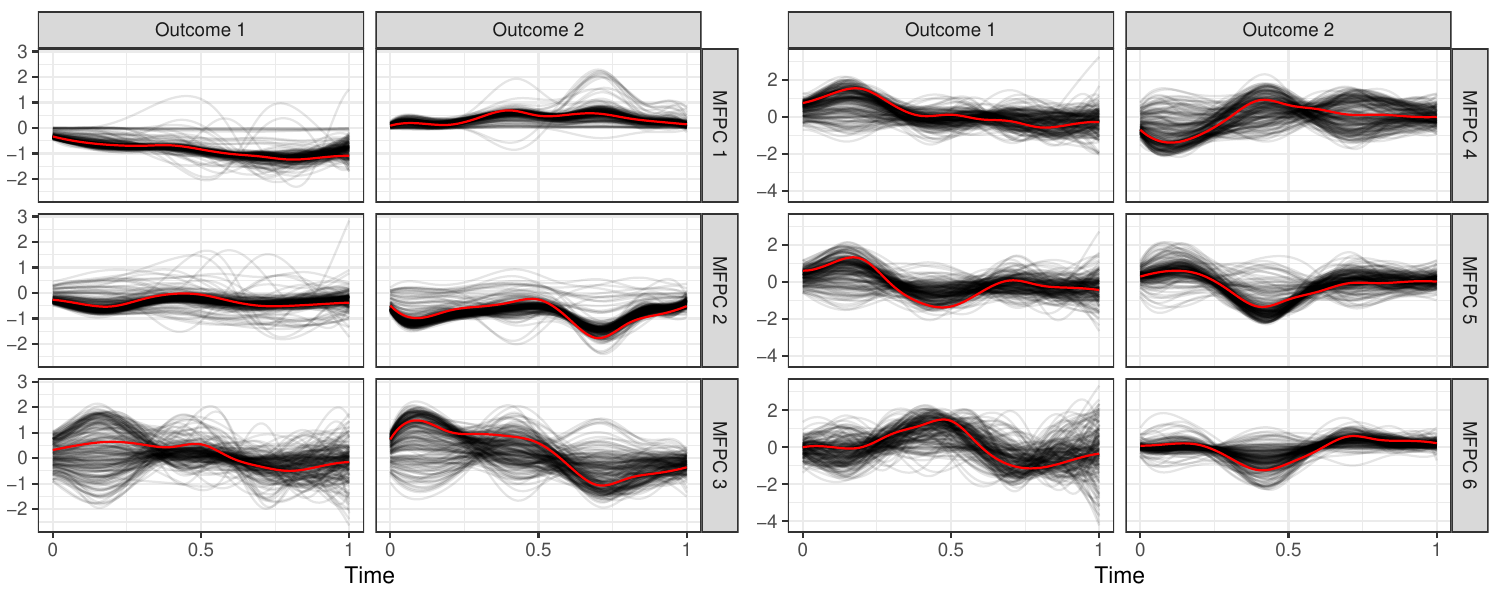}
    \caption{Estimated \gls{mfpc} basis functions $\hat{\bm{\psi}}_m(t)$ for all six eigenfunctions (red) in simulation scenario II. Estimates are flipped if a reflected version $-\hat{\bm{\psi}}_m(t)$ is closer to the true $\bm{\psi}_m(t)$.}
    \label{APPfig:sim_scenII_mfpcs}
\end{figure}

Figure \ref{APPfig:sim_scenI_fits} shows simulated data and model fits for one simulation iteration in scenario I. Figure \ref{APPfig:sim_scenI_mfpcs} shows the estimated \glspl{mfpc} in simulation scenario I, where the true eigenfunctions are linear. Figure \ref{APPfig:sim_scenII_mfpcs} shows the true multivariate eigenfunctions and their estimated counterparts in simulation scenario II. Table \ref{APPtab:mfpc_eval} presents the estimation accuracy of the \glspl{mfpc} in both scenarios as measured by the norm of the differences between true functions and estimates. 

\begin{table}
\caption{Mean multivariate norm of the differences between the eigenfunctions and the estimated \glspl{mfpc} for both simulation scenarios.}
\label{APPtab:mfpc_eval}
\vspace{1em}
\footnotesize
\centering
\begin{tabular}{rrrrrrrrrrrrr}
\hline
 MFPC & 1 & 2 & 3 & 4 & 5 & 6 & 7 & 8 & 9 & 10 & 11 & 12\\
\hline
Scenario I &  0.57 & 1.12 & 1.07 & 1.12 & 0.96 & 0.52 & 0.65 & 0.90 & 1.02 & 1.04 & 1.00 & 0.67 \\ 
 Scenario II &  0.12 & 0.23 & 0.85 & 0.80 & 0.64 & 0.71 & -- & -- & -- & -- & -- & --\\
\hline
\end{tabular}
\end{table}

\section{Application}
\label{app:application}

\begin{figure}
    \centering \includegraphics[width=\textwidth]{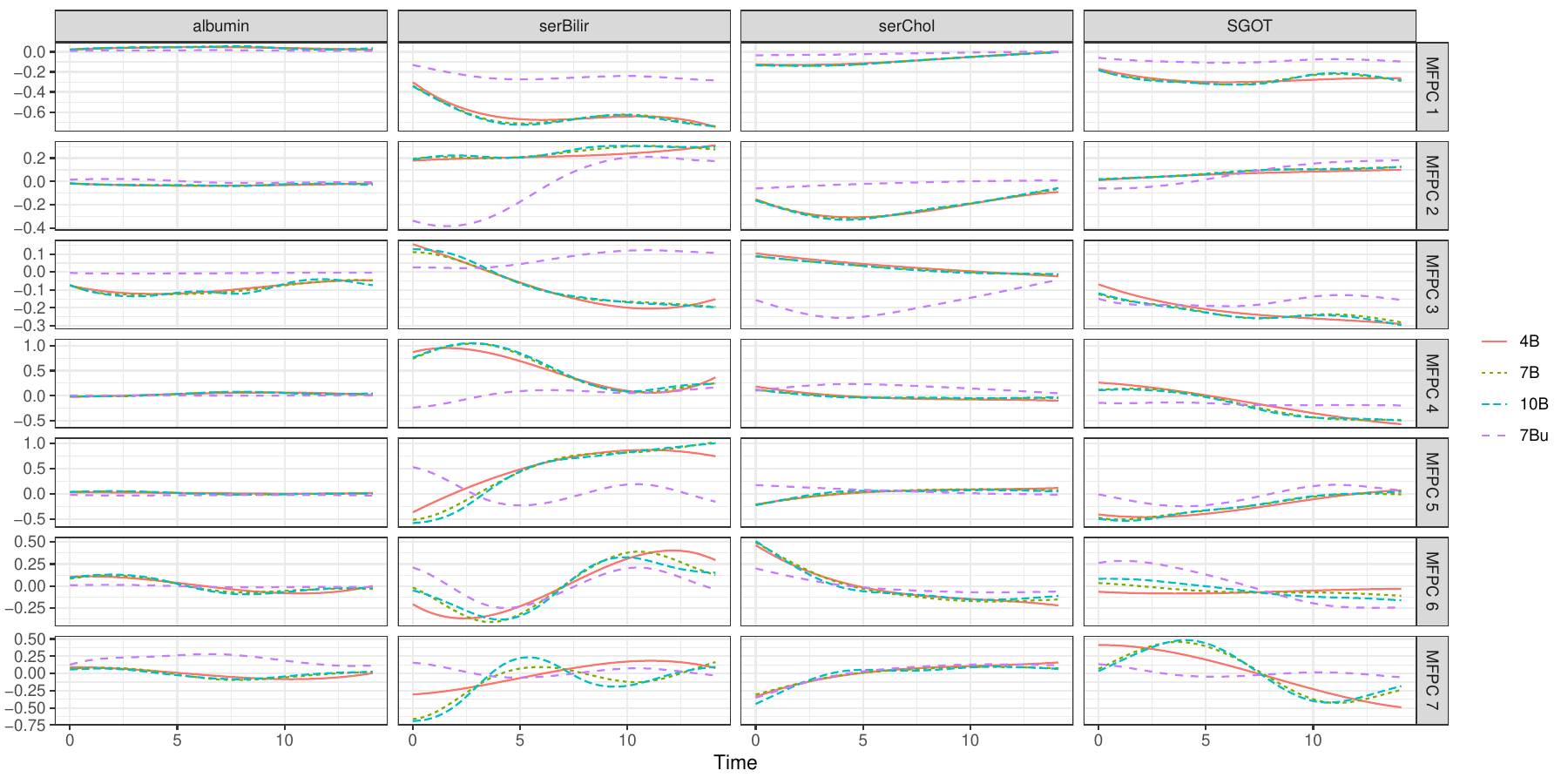}
    \caption{Seven leading estimated \gls{mfpc} basis functions depending on the number of B-spline basis functions used for the estimation of the univariate covariance functions: four (\textit{4B}), seven (\textit{7B}), or ten (\textit{10B}) basis functions. Included are also the corresponding estimates from the unweighted \gls{mfpca} (\textit{7Bu}).}
    \label{APPfig:pbc_mpfcs}
\end{figure}

Figure \ref{APPfig:pbc_mpfcs} shows the seven leading estimated \glspl{mfpc} for different specifications in the univariate covariance estimation and different weightings in the multivariate scalar product. Table \ref{APPtab:pbc_pars} presents the estimated parameters in the \gls{pbc} application for both the \gls{mfpc} based (\textit{bamlss}) and the spline based (\textit{JMB}) approaches.

\begin{table}
\caption{Posterior mean and $95\%$ credible interval of the parameter estimates for the two modeling frameworks.}
\label{APPtab:pbc_pars}
\centering
\vspace{1em}
\footnotesize
\begin{tabular}{l|rr|rr}
\hline
& \multicolumn{2}{c|}{\textit{bamlss}} & \multicolumn{2}{c}{\textit{JMB}}\\
\hline
$\beta_{\gamma1}$ & $-0.20$ & $[-0.71$; $0.31]$ & $-0.18$ & $[-0.67$;  $0.31]$ \\  
$\beta_{\gamma2}$ & $-0.05$ & $[-0.40$; $0.31]$ & $-0.02$ & $[-0.39$;  $0.35]$ \\  
$\beta_{albumin0}$ & $1.20$ & $[1.17$; $1.24]$ & $1.36$ & $[1.28$;  $1.44]$ \\  
$\beta_{albumin1}$ & $-0.03$ & $[-0.07$; $0.00]$ & $-0.04$ & $[-0.08$;  $0.00]$ \\  
$\beta_{albumin2}$ & $0.00$ & $[-0.02$; $0.03]$ & $0.01$ & $[-0.02$;  $0.03]$ \\  $\beta_{serBilir0}$ & $1.00$ & $[0.72$; $1.26]$ & $0.39$ & $[-0.26$;  $1.01]$ \\  $\beta_{serBilir1}$ & $-0.04$ & $[-0.31$; $0.23]$ & $0.00$ & $[-0.30$;  $0.29]$ \\  $\beta_{serBilir2}$ & $-0.02$ & $[-0.22$; $0.17]$ & $-0.06$ & $[-0.25$;  $0.14]$ \\  $\beta_{serChol0}$ & $5.61$ & $[5.50$; $5.72]$ & $5.84$ & $[5.58$;  $6.10]$ \\  $\beta_{serChol1}$ & $0.06$ & $[-0.05$; $0.17]$ & $0.07$ & $[-0.05$;  $0.19]$ \\  $\beta_{serChol2}$ & $0.05$ & $[-0.02$; $0.12]$ & $0.04$ & $[-0.04$;  $0.11]$ \\  $\beta_{SGOT0}$ & $4.75$ & $[4.59$; $4.90]$ & $4.75$ & $[4.46$;  $5.04]$ \\  $\beta_{SGOT1}$ & $0.02$ & $[-0.13$; $0.18]$ & $-0.02$ & $[-0.15$;  $0.12]$ \\  $\beta_{SGOT2}$ & $-0.06$ & $[-0.15$; $0.05]$ & $-0.10$ & $[-0.19$;  $-0.01]$ \\   $\beta_{\sigma albumin}$ & $-2.29$ & $[-2.33$; $-2.26]$ & $-2.33$ & $[-2.37$;  $-2.30]$ \\  $\beta_{\sigma serBilir}$ & $-1.19$ & $[-1.23$; $-1.15]$ & $-1.21$ & $[-1.25$;  $-1.17]$ \\  $\beta_{\sigma serChol}$ & $-1.71$ & $[-1.76$; $-1.65]$ & $-1.73$ & $[-1.79$;  $-1.67]$ \\  $\beta_{\sigma SGOT}$ & $-1.33$ & $[-1.37$; $-1.29]$ & $-1.34$ & $[-1.38$;  $-1.30]$ \\ 
\hline
\end{tabular}
\end{table}

\end{document}